\documentclass[fleqn,usenatbib]{mnras}

\usepackage{amssymb}	% Extra maths symbols
\usepackage{booktabs,tabularx,graphicx,amsmath,makecell,hhline,enumitem,gensymb,multirow,ulem,pifont,adjustbox}
\usepackage{hyperref} % links
\usepackage[dvipsnames]{xcolor}
\usepackage{newtxtext,newtxmath}
\usepackage[T1]{fontenc}

\DeclareRobustCommand{\VAN}[3]{#2}
\let\VANthebibliography\thebibliography
\def\thebibliography{\DeclareRobustCommand{\VAN}[3]{##3}\VANthebibliography}

\usepackage{graphicx}	% Including figure files
\usepackage{amsmath}	% Advanced maths commands
\usepackage{enumitem}
\newcommand{\mycomment}[1]{}

\title[Radially anisotropic Milky Way-like systems]{Exploring the diversity and similarity of radially anisotropic Milky Way-like
stellar haloes: implications for disrupted dwarf galaxy searches}

\author[M. D. A. Orkney et al.]{Matthew D. A. Orkney,$^{1,2}$\thanks{E-mail: morkney@icc.ub.edu}
Chervin F. P. Laporte,$^{1,2}$ Robert J. J. Grand,$^{3,4,5}$ Facundo A. G\'omez,$^{6}$ \newauthor Freeke van de Voort,$^{7}$ Azadeh Fattahi,$^{8}$ Federico Marinacci,$^{9}$ R\"udiger Pakmor,$^{10}$ Francesca Fragkoudi,$^{8}$ \newauthor and Volker Springel$^{10}$\vspace{0.1cm}\\
$^{1}$Institut de Ci\`{e}ncies del Cosmos (ICCUB), Universitat de Barcelona, Mart\'{i} i Franqu\`{e}s 1, E-08028 Barcelona, Spain\\
$^{2}$Institut d'Estudis Espacials de Catalunya (IEEC), E-08034 Barcelona, Spain\\
$^{3}$Astrophysics Research Institute, Liverpool John Moores University, 146 Brownlow Hill, Liverpool, L3 5RF, UK\\
$^{4}$Instituto de Astrof\'isica de Canarias,
Calle Via Lactea s/n, E-38205 La Laguna, Tenerife, Spain\\
$^{5}$Departamento de Astrof\'isica, Universidad de La Laguna,
Av. del Astrof\'isico Francisco S\'anchez s/n, E-38206, La Laguna, Tenerife, Spain\\
$^{6}$Departamento de Astronom\'ia, Universidad de La Serena, Av. Juan Cisternas 1200 Norte, La Serena, Chile\\
$^{7}$Cardiff Hub for for Astrophysics Research and Technology, School of Physics and Astronomy, Cardiff University, Queen's Buildings, The Parade, Cardiff\\ CF24 3AA, UK \\
$^{8}$Institute for Computational Cosmology, Department of Physics, Durham University, South Road, Durham DH1 3LE, UK \\
$^{9}$Department of Physics \& Astronomy ``Augusto Righi'', University of Bologna, via Gobetti 93/2, 40129 Bologna, Italy\\
$^{10}$Max Planck Institut f{\"u}r Astrophysik, Karl-Schwarzschild-Straße 1, 85748 Garching bei München, Germany \\
}

\date{Accepted 08 August 2023.}

\pubyear{2023}

\begin{document}
\label{firstpage}
\pagerange{\pageref{firstpage}--\pageref{lastpage}}
\maketitle

\begin{abstract}
%
% Bordering 250 words.
%
We investigate the properties of mergers comparable to the Gaia-Sausage-Enceladus (GSE) using cosmological hydrodynamical simulations of Milky Way-like galaxies. The merger progenitors span an order of magnitude in their peak stellar mass ($3\times10^8<M_{\star}/\rm{M}_{\odot}<4\times10^9$) and include both rotation and pressure-supported galaxies ($0.10<D/T<0.77$).
In a minority of cases, the GSE-like debris is comprised of stars from more than one merger progenitor. However, there is a close similarity in their chemodynamical properties and the triaxial shapes of their debris, and so it is not always possible to distinguish them. 
The merger progenitors host a variety of luminous satellites ($0\mathit{\textnormal{-}}8$ with $M_{\star}>10^6\,\rm{M}_{\odot}$), but most of these do not follow the merger to low orbital energies. Between $0\mathit{\textnormal{-}}1$ of these satellites may survive to $z=0$, but with no clear signatures of their past association.
We show that the fraction of stars originating from GSE-like mergers is reduced for lower metallicities (reaching a minimum around $\text{[Fe/H]} = -2$), and also within 5\,kpc of the galactic centre. Whilst these central regions are dominated by \textit{in-situ} stars, the \textit{ex-situ} fraction trends towards a 100 per cent asymptote when considering the most metal-poor stars ($\text{[Fe/H]}\ll-2.5$). Considering this, its near proximity, and its small volume on the sky, the Galactic centre lends itself as a prime environment in the search for the stars from the earliest galaxies, whilst avoiding contamination from GSE stars.
\end{abstract}


%%%%%%%%%%%%%%%%%%%%%%%%%%%%%%%%%%%%%%%%%%%%%%%%%%

%%%%%%%%%%%%%%%%% BODY OF PAPER %%%%%%%%%%%%%%%%%%

\section{Introduction}

In the Lambda Cold Dark Matter ($\Lambda$CDM) paradigm, present-day galaxies are formed through the hierarchical assembly of numerous mergers \citep{press1974, white1978, searle1978, blumenthal1984, white1991, kauffmann1993}, and their final properties are a product of this intricate assembly process. Galaxy mergers often produce visible substructure within the stellar haloes of the host galaxy, such as surviving remnant objects \citep[e.g. the Sagittarius dwarf in the MW][]{ibata1994, majewski2003}, shells \citep[e.g.][]{schweizer1986, martinez2008, arjun2023}, or streams \citep[e.g.][]{bell1995, belokurov2006, belokurov2007, martinez2015}. However, many of the most ancient accreting galaxies will have gradually dissolved due to cumulative perturbations and dynamical relaxation \citep[see][]{ibata2002, mayer2002, pen2006, gomez2013, maffione2015, buist2015, whelan2016, ngan2016, erkal2016, maffione2018}, and eventually become smoothly integrated within the stellar halo. In these cases, there may yet be evidence of their existence in the form of lingering features in the chemical and dynamical planes. \par

Observations infer that ancient Milky Way (MW) type galaxies underwent a phase of rapid merging in the early Universe \citep{unavane1996, bell2008}, and this is reinforced by the predictions of cosmological simulations \citep{bullock2005, bell2008, cooper2010, Kruijssen2019, renaud2021}. Fortunately, the progenitor galaxies of these mergers can be distinguished by their unique chemical evolution, which is itself tied to the details of their formation history \citep[e.g.][]{freeman2002, venn2004, gallart2005, robertson2005, tolstoy2009}. Furthermore, their debris is linked by shared integrals of motion that relate to their infall and internal kinematics \citep[e.g.][]{helmi2000, font2011, amarante2022}. Given the long dynamical times within the stellar halo of MW-mass galaxies ($\mathcal{O}(\rm{Gyrs})$), chemodynamic signatures of ancient mergers may persist to the present day \citep[e.g.][]{eggen1962, freeman2002, helmi2003, nissen2010, hayes2018, vera2022}, albeit with some degree of phase-mixing. Investigating these signatures, and using them to reconstruct the history of our own Galaxy, is a core goal in the field of Galactic Archaeology. \par

% History of the GSE:
%%%%%

Early exploration of old metal-poor halo stars around the Solar neighbourhood revealed a characteristic radially-anisotropic velocity distribution \citep{chiba2000}. This result was interpreted in the context of preexisting formation models of the inner-galaxy: that this population of stars formed from the monolithic collapse of radially infalling gasses \citep{eggen1962}, and/or they were inherited from radially infalling dwarf galaxy mergers \citep{searle1978, helmi1999}, or even that they were thick-disc stars which were dynamically heated onto radial orbits following a disruptive merger event \citep[e.g.][]{quinn1993, villalobos2008}. Chemodynamical simulations favoured the satellite accretion scenario \citep{brook2003}, with some citing observed $\alpha$-abundance versus metallicity trends as evidence for a relation to the hypothetical progenitor galaxy of the $\omega$ Cen globular cluster \citep{meza2005, nissen2010, limberg2022}. \par

Astrometric and photometric data from the \textit{Gaia} \citep{GaiaDR1} and SDSS \citep{sdss52017} missions have supported more extensive explorations of the Solar neighbourhood than ever before. This has revealed that the velocities of MW halo stars are highly radially anisotropic at sub-Solar metallicities \citep{belokurov2018}. The velocity anisotropy can be parameterised following \citet{binney1980} as:
\begin{equation}
\beta(r) = 1 - \frac{\left \langle v_t^2 \right \rangle}{2\left \langle v_r^2 \right \rangle},
\label{equ:anisotropy}
\end{equation}
where $\left \langle v_t^2 \right \rangle$ and $\left \langle v_r^2 \right \rangle$ are the squared velocity dispersions in the tangential and radial directions. Here, $\beta = 1$ describes a velocity distribution that is radially anisotropic, $\beta=0$ is isotropic, and $\beta=-\infty$ is tangentially anisotropic. Following this definition, the anisotropy parameter of MW halo stars peaks at $\beta \sim 0.9$ for metallicities $\text{[Fe/H]} > -1.7$ \citep{belokurov2018}. This extreme value cannot easily be explained by an accretion of numerous dwarf galaxies, because the orbital variation between each progenitor galaxy would yield a more isotropic velocity distribution in their debris. Instead, \citep{belokurov2018} invoked a single massive dwarf accretion at a redshift between $1<z<3$ with virial mass $>10^{10}\,\rm{M}_{\odot}$.
This scenario was corroborated elsewhere due to discoveries of breaks in the stellar halo density profile \citep{deason2013, amorisco2017}, evidence for a two-component halo in colour-magnitude diagrams \citep{gaia2018, haywood2018, helmi2018}, stellar kinematics \citep{koppelman2018}, and stellar ages \citep{gallart2019}. In particular, \citet{helmi2018} showed that the the $\alpha$-abundance versus metallicity trends of these halo stars were more consistent with having formed in a separate galaxy than in the MW thick disc. This possible accretion event has since become known as the Gaia-Sausage-Enceladus (GSE), and evidence favouring its existence has continued to accumulate \citep[e.g.][]{mackereth2019, bignone2019, naidu2020, myeong2022}. However, constraining various progenitor properties, including the absolute number of progenitors, may be more difficult than anticipated \citep[i.e. see][]{rey2023}. \par

As 6D phase-space and chemical observations of the stellar halo have improved, it has become possible to identify substructures related to the GSE debris \citep{simon2019, perottoni2022}, as well as a myriad of other possibly distinct debris features (e.g. \textit{Sequoia:} \citealp{myeong2018, barba2019, myeong2019}, \textit{Kraken:}\footnote{Also known as \textit{Heracles}.} \citealp{Kruijssen2019, horta2021, naidu2022}, \textit{Thamnos:} \citealp{koppelman2019}, \textit{Wukong:} \citealp{yuan2020, naidu2020}, \textit{Icarus} \citep{fiorentin2021}, \textit{Pontus:} \citealp{malhan2022}, etc.). Whilst evidence in favour of the GSE massive merger scenario has continued to grow, the consensus around the origins of these other debris groups is more uncertain. The very retrograde high-energy halo debris may be the aggregate of three independent accretion events \citep{myeong2019, naidu2020} or stars from the outskirts of the GSE progenitor \citep{koppelman2020, naidu2021, amarante2022}. There are arguments that \textit{Kraken} could be a population of low-metallicity \textit{in-situ} stars \citep{belokurov2022, myeong2022, rix2022, orkney2022}, though statistical methods in \citet{horta2022} suggest that \textit{Kraken} is distinguishable from this \textit{in-situ} population by its lower $\alpha$-abundance. Some other objects are difficult to differentiate from the pervasive GSE debris, and may represent overdensities in a fragmented GSE debris footprint \citep[e.g.][]{amarante2022}, or possibly satellite galaxies of other major progenitors. Furthermore, the tendency for different merger debris distributions to overlap and fragment can make it impossible to correctly identify their origins \citep{jeanbaptiste2017}. A more rigorous consideration of the GSE debris, its extents in various chemical and dynamical properties, and its contribution fraction compared to other debris groups, would help to substantiate or relieve these concerns. \par

% This paper:
%%%%%

In this paper, we analyse a selection of MW-type galaxies from the {\sc auriga} simulation suite. \citet{fattahi2019}, hereafter F19, show that these galaxies are host to radially anisotropic inner-halo debris features with properties broadly comparable to the GSE, and further identify the main mergers that contribute to these debris features. We explore the composition of these debris features, finding in some cases there are two or three separate mergers that contribute non-negligible mass fractions. We then investigate the properties of the merger that contributes the largest fraction of the GSE-like debris, both in terms of their pre-infall galaxies and the properties of their resulting debris at $z=0$. These mergers span a wide range of properties, but these are not necessarily a good predictor of the properties in their debris. \par

We describe the {\sc auriga} suite, our simulation sample and our post-processing choices in Section \ref{sec:method}. Our results are shown in Section \ref{sec:results}: we illustrate the contributions of each merger to their respective GSE-like features in Section \ref{sec:debris_fractions}, then we investigate a selection of different properties for both the progenitor mergers and their debris in Section \ref{sec:prog}, their populations of luminous satellites in Section \ref{sec:sats}, and overall trends in the halo debris in Section \ref{sec:debristrends}. We discuss the implication of these results in Section \ref{sec:discussion}, and conclude in Section \ref{sec:conclusion}. \par

\section{Method} \label{sec:method}

% Describe the Auriga simulation suite:

\subsection{Simulation suite}

The simulations analysed in this paper are taken from the {\sc auriga} project \citep{Auriga}. {\sc Auriga} includes thirty magneto-hydrodynamic simulations of isolated MW-mass galaxies within a full cosmological context. These galaxies represent a wide diversity of accretion histories, and were not specifically designed to match the assembly history of the MW. \par

The target galaxies were originally selected from the Dark Matter (DM) only version of the Ref-L100N1504 cosmological volume in the {\sc eagle} project \citep{Eagle}, and then resimulated using the Tree-PM moving-mesh code {\sc arepo} \citep{Arepo} using the `zoom' approach \citep{katz1993, frenk1996}. The initial conditions were generated with the Gaussian white-noise realisation {\sc panphasia} \citep{Panphasia}, within a periodic cosmological box of side length $100\,$cMpc. The suite uses cosmological parameters from \citet{Planck2014}, which are $\Omega_{\rm m}=0.307$, $\Omega_{\rm b}=0.04825$, $\Omega_{\rm \Lambda}=0.693$ and a Hubble constant of $H_0=100h\,\text{km}^{-1}\,\text{Mpc}^{-1}$, where $h=0.6777$. \par

{\sc Auriga} includes physical models for a spatially uniform photoionizing UV background, primordial and metal line cooling, star formation, stellar evolution and supernovae feedback, supermassive black hole growth and feedback, and magnetic fields. See \citet{Auriga} for a description of these sub-grid physics models, which are able to reproduce a range of expected galaxy properties in cosmological contexts \citep{vogelsberger2013, marinacci2014, genel2014}. The {\sc auriga} galaxies have realistic properties that are generally compatible with MW-type galaxies in terms of their halo mass-metallicity relations \citep{monachesi2019}, disc oscillations \citep{gomez2017}, rotation curves and star formation rates \citep{Auriga}, thin/thick disc dichotomy \citep{grand2018}, and the chemodynamical properties of their innermost regions \citep{fragkoudi2020}. \par

In this work, we utilise the ``level-4'' resolution versions of each {\sc auriga} simulation, where the target high-resolution region is resolved with a DM particle mass of $\sim3\times10^5\,\text{M}_{\odot}$ and baryonic mass of $\sim5\times10^4\,\text{M}_{\odot}$. Each simulation includes 128 snapshots over the range $127 \geqslant z \geqslant 0$. Hereafter we refer to the {\sc auriga} simulations as Au-$i$, where $i$ indicates the particular halo. \par

\subsection{Post-processing} \label{sec:post}

Virial properties and a varied assortment of group and subhalo properties are calculated using the {\sc subfind} halo finder \citep{subfind}, and are derived using a sphere of mean density $200\times$ the critical density of the universe. Haloes and subhaloes are linked across timesteps using the {\sc lhalotree} merger tree algorithm \citep{springel2005}. \par

{\sc Auriga} tracks a selection of different chemical abundances, including iron and $\alpha$-process elements. We normalise the abundance ratios to Solar values following \citet{asplund2009}, and apply further systematic shifts of $-0.4$ to the [Fe/H] ratios (and equivalent for other abundance ratios). Similar corrections were also used in F19 and \citet{grand2020}, and are calibrated based on observations of the MW abundance ratios. \par

We designate a star particle as `\textit{in-situ}' if it was bound to the potential well of the main progenitor halo at its formation time (as in \citealt{cooper2015}), and `\textit{ex-situ}' if it formed within the potential well of satellite haloes or their progenitors. This choice means that stars forming within recently stripped gas are counted as \textit{in-situ}, although realistically their identification could be more ambiguous. Therefore, we expect this method to represent an upper-bound on the fraction of \textit{in-situ} stars. \par

All galaxies are reoriented to align on the angular momentum of the innermost \textit{in-situ} star particles ($R_{\rm G}<0.1\times R_{200}$, where $R_{\rm G}$ is the galacto-centric radius), such that any co-rotating discs are viewed face-on in the $x$-$y$ plane. The gravitational potential energies of each star particle are recovered directly from the simulation data, and are normalised such that the spherically-symmetric potential profile at the $R_{200}$ virial radius is 0. \par

It is occasionally necessary to distinguish the co-rotating stellar disc from the stellar halo. We achieve this by calculating the circularity parameter for each star particle (as in \citealt{abadi2003, Auriga, gomez2017b}), defined as:
\begin{equation}
\epsilon = \frac{L_{z}}{L_{z}^{\rm max}(E)},
\label{equ:e_circ}
\end{equation}
where $L_{z}$ is the $z$-component of the angular momentum, and $L_{z}^{\rm max}(E)$ is the maximum angular momentum that is allowed for the given specific orbital energy. Here, $\epsilon=1$ corresponds to prograde circular orbits in the plane of the disc, $\epsilon=-1$ corresponds to retrograde orbits, and $\epsilon=0$ corresponds to orbits with low angular momentum in the $z$ direction. \par

We follow the assumption that each galaxy consists of a stellar spheroid with net zero rotational velocity and a symmetric distribution centered on $\epsilon=0$ (the stellar halo), and a component that is co-rotating with $\epsilon>0$ (the stellar disc/bar). The spheroid is estimated by mirroring the retrograde $\epsilon$ distribution about 0. The probability of a star particle being in the halo ($p_{\rm halo}$) can be estimated by dividing the total $\epsilon$ distribution by the spheroidal $\epsilon$ distribution. Each star particle is then assigned to the disc or halo by drawing a random number $n$ in the range $0\mathit{\textnormal{-}}1$, where $n>p_{\rm halo}$ corresponds to a disc star. \par

This method does not discriminate between different kinds of co-rotating distributions, such as stellar discs, bars or clumpy features. The probabilistic determination used in this technique means there is always a chance that a disc particle is erroneously assigned to the high-$\epsilon$ portion of the halo, and vice versa. The disc and halo populations should typically have unique chemical abundance distributions, with disc stars tending to be more metal-rich than halo stars, and these distributions will become smeared together in our selection. Therefore, we include the additional requirement that the retrograde and prograde sides of the stellar halo should share the same [Fe/H] distribution function. This assumption is incorporated into our calculation of $p_{\rm halo}$, but we note that none of our reported results are dependent on it. \par

\subsection{GSE-like features in Auriga}

\begin{table*}
\setlength{\tabcolsep}{4pt} % default is 6pt
\resizebox{\textwidth}{!}{
\begin{tabular}{lcc|cc|cc|ccc} 
\toprule
Run & GSE-type merger & $z_{\rm infall}$ & \makecell{$M_{\rm 200}$(pre-infall) \\ $[10^{10}\,\rm{M}_{\odot}]$} & \makecell{$M_{200}/M_{\rm 200,Host}$ \\ (pre-infall)} & \makecell{$M_{\rm gas}$(pre-infall) \\ $[10^{10}\,\rm{M}_{\odot}]$} & \makecell{$f_{\rm gas}$ \\ (pre-infall)} & \makecell{$M_{\star}$(pre-infall/peak) \\ $[10^{10}\,\rm{M}_{\odot}]$} & \makecell{$R_{\rm half, \star}$(pre-infall/peak) \\ $[\rm{kpc}]$} & \makecell{$D/T$ \\ (pre-infall)} \\
\midrule
Au-5 & Au-5-M & 0.90 & 9.59 & 0.24 & 1.73 & 0.53 & 0.32 / 0.38 & 2.51 / 2.63 & 0.10 \\
Au-9 & Au-9-M & 1.91 & 7.02 & 0.68 & 1.43 & 0.84 & 0.15 / 0.19 & 3.66 / 5.24 & 0.30 \\
Au-10 & Au-10-M1 & 0.90 & 3.62 & 0.08 & 0.88 & 0.71 & 0.09 / 0.10 & 4.40 / 4.41 & 0.77 \\
      & Au-10-M2 & 0.75 & 3.33 & 0.06 & 0.73 & 0.63 & 0.08 / 0.09 & 2.24 / 2.43 & 0.18 \\
Au-15 & Au-15-M & 0.82 & 10.92 & 0.41 & 2.08 & 0.44 & 0.22 / 0.25 & 4.49 / 3.84 & 0.57 \\
Au-17 & Au-17-M & 2.47 & 3.12 & 0.11 & 0.70 & 0.77 & 0.03 / 0.04 & 1.86 / 2.21 & 0.20 \\
Au-18 & Au-18-M & 1.39 & 3.82 & 0.16 & 0.87 & 0.64 & 0.13 / 0.14 & 3.10 / 3.16 & 0.10 \\
Au-22 & Au-22-M & 2.85 & 2.54 & 0.20 & 0.62 & 0.83 & 0.03 / 0.03 & 2.29 / 2.27 & 0.21 \\
Au-24 & Au-24-M & 1.35 & 9.13 & 0.15 & 1.65 & 0.42 & 0.22 / 0.26 & 3.72 / 3.56 & 0.48 \\
Au-27 & Au-27-M & 1.59 & 10.82 & 0.25 & 1.94 & 0.64 & 0.39 / 0.41 & 3.89 / 3.98 & 0.57 \\
\bottomrule
\end{tabular}
}
\caption{The properties of the GSE-type mergers considered in this work. Here, $z_{\rm infall}$ is the time at which the merger passes the coeval $R_{200}$ virial radius of the central {\sc auriga} galaxy. Pre-infall masses are determined for all member particles at the snapshot prior to $z_{\rm infall}$, and are therefore minimally impacted by gravitational tides. The merger mass fraction ($M_{200}/M_{200\rm, Host}$) is found at the last snapshot where each object is a distinct group according to {\sc subfind}. Gas fractions are calculated as $M_{\rm gas}/(M_{\rm gas} + M_\star)$, for all material within twice the stellar half mass radius. We also include the peak stellar masses using all bound stars found by {\sc subfind}, and corresponding stellar half mass radii. The pre-infall disc to total mass ratio, as defined in Section \ref{sec:post}, is given as $D/T$.}
\label{tab:gse_analogue_properties}
\end{table*}

F19 investigated the velocity anisotropy of inner-halo stars for 28 {\sc auriga} simulations. They identified MW-like galaxies in 10 simulations which possessed prominent radially anisotropic debris features with $\beta > 0.8$ and contributions to the stellar halo greater than 50 per cent, which were considered to be comparable to the GSE.
These simulations were Au-5, 9, 10, 15, 17, 18, 22, 24, 26 \& 27. \par

F19 also identified the progenitor merger that contributed the greatest mass fraction of this debris feature within specific metallicity, galactic height, and velocity ranges. In many cases these mergers are responsible for the majority of the mass fraction, but in some other cases they are responsible for only a plurality. We focus on only the main progenitor mergers for the purposes of this paper, and ignore any other mergers that also contribute to the radially anisotropic feature. We make an exception for Au-10, for which there are two mergers that contribute almost equivalent mass fractions at the Solar radius, and in this case we investigate both objects to see if they can be distinguished. We exclude Au-26 from our analysis because the main progenitor merger is so massive ($M_\star>10^{10}\,\rm{M}_{\odot}$) that it is difficult to reconcile with the expected properties of the GSE. We list these mergers, alongside a selection of key properties, in Table \ref{tab:gse_analogue_properties}. \par

Throughout this paper, we use the nomenclature `GSE-type' to refer to the merger identified as contributing the most stars to the radially-anisotropic debris feature. Similarly, `GSE-like debris' refers only to the debris originating from this particular merger event. We refer to the MW-mass galaxy as the `central {\sc auriga}' galaxy. \par

\section{Results} \label{sec:results}

\subsection{Radially anisotropic debris} \label{sec:debris_fractions}

\begin{figure*}
\centering
  \setlength\tabcolsep{2pt}%
    \includegraphics[keepaspectratio, trim={0.0cm 0.0cm 0.0cm 0.5cm}, width=0.8\linewidth]{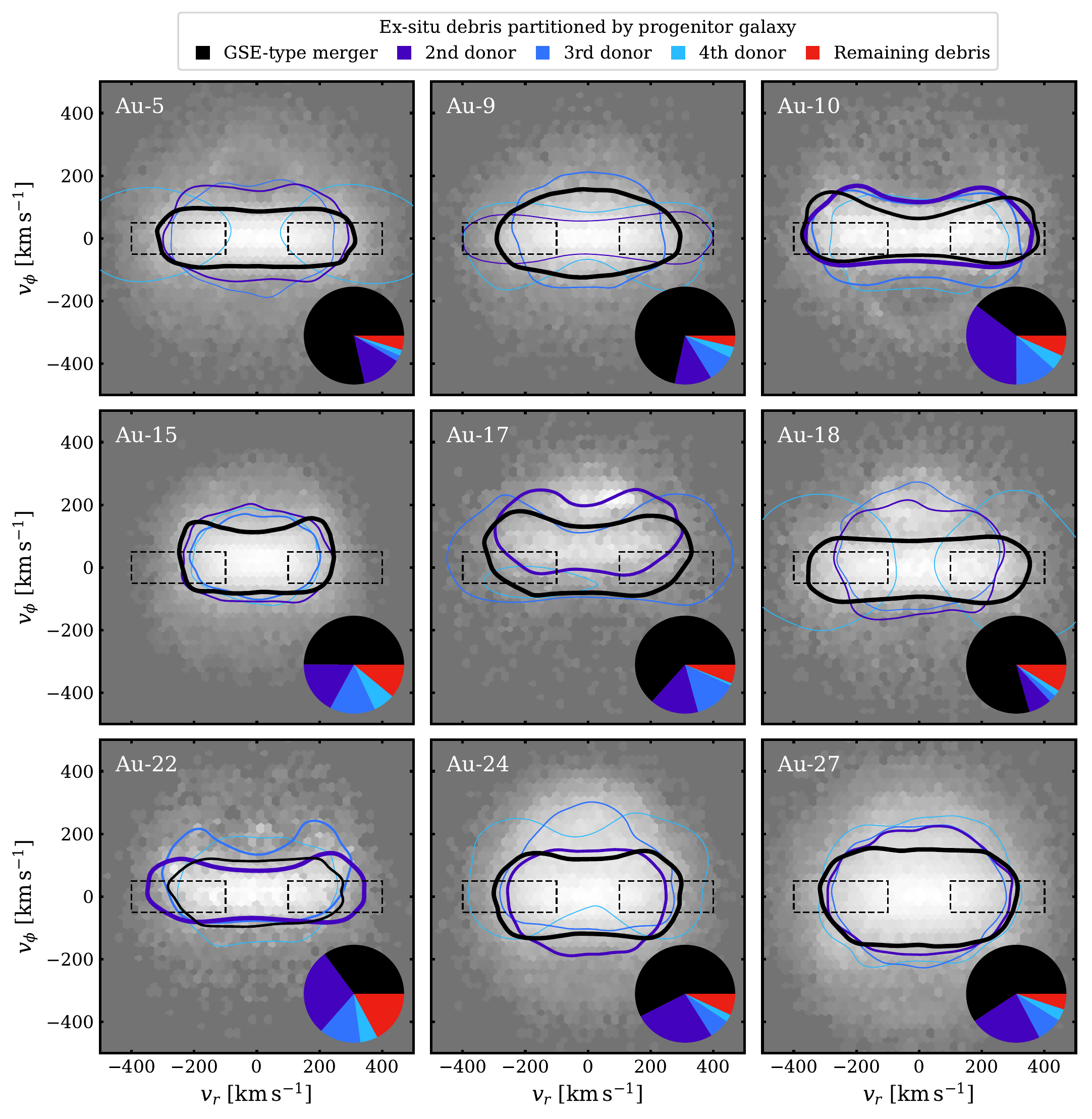}\\
\caption{Each panel shows a histogram of \textit{ex-situ} stellar mass in $v_{\phi}/v_{\rm r}$ spherical polar coordinates, with the following parameter cuts: $R_{\rm G}<30\,$kpc; $-2<\text{[Fe/H]}<-0.5$; $|Z|>3\,$kpc. The pie charts in the lower-right hand corners show the mass contribution fractions of stars within the radially anisotropic black-dashed rectangle regions: $|v_{\phi}|<50\,\rm{km}\,\rm{s}^{-1}$; $100<|v_{\rm r}|/\rm{km}\,\rm{s}^{-1}<400$. Each pie segment represents a different progenitor galaxy as indicated in the legend. The stellar debris from each of these progenitors are indicated with a coloured outline that encircles 80 per cent of their mass, produced using an iso-density contour around a smoothed Gaussian kernel. Line width indicates the total debris mass. These provide a visualisation of the radial velocity elongation. Whilst there is typically one dominant radially anisotropic merger, some realisations have significant contributions from other mergers.}
\label{fig:paper_pie}
\end{figure*}

\begin{figure*}
\centering
  \setlength\tabcolsep{2pt}%
    \includegraphics[keepaspectratio, trim={0.0cm 0.0cm 0.0cm 0.0cm}, width=\linewidth]{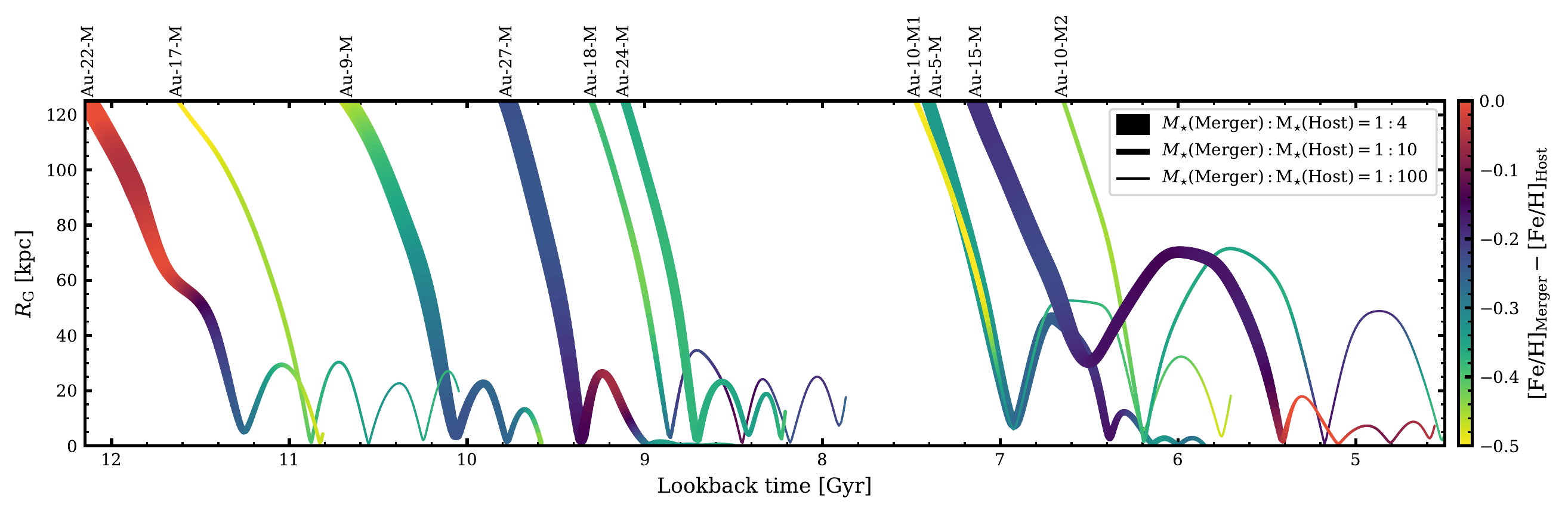}\\
\caption{The infalls of the GSE-type mergers listed in Table \ref{tab:gse_analogue_properties}, as marked on the top-axis. The stellar mass ratio of the merger is represented by the line width, and the difference in mean [Fe/H] between the merging galaxy and central {\sc auriga} galaxy is represented by the line colour (as calculated for stars within the stellar half mass radius). The paths have been fitted with a cubic spline to improve the time resolution, and to better resolve the orbits. The infalls are highly radial, and in most cases the first or second apocentres are within 30\,kpc.}
\label{fig:paper_infall}
\end{figure*}

\subsubsection{\textit{ex-situ} contributions} \label{sec:contributions}

The \textit{ex-situ} stellar density for each selected {\sc auriga} simulation is shown at $z=0$ in Figure \ref{fig:paper_pie} as a series of grey-scale histograms. We perform a cut on stars with galacto-centric radii $R_{\rm G}<30\,$kpc, metallicities $-2<\text{[Fe/H]}<-0.5$, and height above the disc plane of $|Z|>3\,$kpc, which is intended to generously bracket the expected properties of GSE-type debris. \par

We define a ``radially anisotropic debris'' region to be all \textit{ex-situ} stars with polar velocity coordinates of $|v_{\phi}|<50\,\rm{km}\,\rm{s}^{-1}$ and $100<|v_{\rm r}|/\rm{km}\,\rm{s}^{-1}<400$, as indicated by the dashed black rectangles in each panel. These are similar to the kinematic cuts employed in F19, and are designed to eliminate most of the isotropic halo stars. We include a pie chart in the lower-right corner of each panel, which displays the mass contribution fractions within these velocity cuts. The stars from the four most major contributors are shown with unique colours (black, dark-blue, blue, light-blue), and the debris from all remaining progenitor galaxies is shown in combination (red). \par

We take the four mergers that contribute the greatest radially-anisotropic mass fractions, and draw contours around the perimeters of their debris in order to convey the shapes of their full velocity distributions. This is based upon an iso-density contour of a smoothed Gaussian kernel density estimate that encircles 80 per cent of the stellar mass, or else a mass-weighted area density of $10\,\rm{M}_{\odot}\,\rm{km}^{-2}\,\rm{s}^{2}$ if the star particles are sparsely distributed. The contour line width reflects the total mass contribution without the velocity cuts, and is therefore complementary to the pie chart. The debris from the GSE-type merger is in all cases radially anisotropic, with $\beta>0.6$ over the property ranges used here. \par

The debris from other significant mergers adopt a range of contribution fractions and anisotropy. In Au-9 there is a subdominant merger that is even more radially biased ($\beta=0.92$) than the GSE-type merger ($\beta=0.63$), although it contributes only 11 per cent of the anisotropic mass fraction. In Au-10 and Au-22 there are subdominant mergers which contribute high mass fractions and high anisotropy ($\beta>0.7$). In particular, the two greatest contributors in Au-10 are near-equal over a wide range of radii. On the other hand, there are subdominant mergers such as that in Au-24 which has a relatively high mass fraction of 26 per cent but a significantly lower anisotropy ($\beta=0.44$). \par

We show an example merger tree visualisation in Appendix \ref{appendix:mergertrees}, with the evolutionary tracks of the four largest contributors from Figure \ref{fig:paper_pie} in the same colours. We take this opportunity to highlight a problem in the {\sc lhalotree} algorithm used to generate these merger trees. The subhalo identification occasionally stops tracking a merger remnant as it nears a pericentre passage. This behaviour is normally corrected by allowing a halo descendant to skip a snapshot \citep[see][]{springel2005}, but in some cases the remnant is re-identified after emerging on the other side of its pericentre as an entirely independent merger event. We have found two cases where this situation affects the mergers identified in Figure \ref{fig:paper_pie}, and we have corrected this by summing their debris together. As a result of this change, the most massive contributor in Au-22 is a different object to the one found in F19. \par

\subsubsection{Progenitor mass} \label{sec:masses}

The progenitor masses shown in Table \ref{tab:gse_analogue_properties} bracket almost an order of magnitude, and there is a wide variation in their merger mass ratios (from $<1:10$ to $1:4$). Nonetheless, all subsequent mergers have a lower merger mass ratio than these GSE-type mergers.  Furthermore, there are only two examples (Au-15, Au-24) in which there is a subsequent merger with a mass ratio $>1:10$, and only one of these examples (Au-24) contributes a significant stellar mass fraction to the Solar neighbourhood. Therefore, these GSE-type mergers can be considered the last `significant' merger in their respective host galaxies, similar to what is expected for the MW. \par

\citet{belokurov2018} originally estimated that the GSE progenitor would need a virial mass of $>10^{10}\,\rm{M}_{\odot}$, which is consistent with the GSE-type mergers in Table \ref{tab:gse_analogue_properties}. Using the redshift-dependent stellar to halo mass relation of \citet{moster2013}, this would suggest a progenitor stellar mass of $>4\times10^{6}\,\rm{M}_{\odot}$. There have been many more estimates of the GSE progenitor stellar mass, including $5\mathit{\textnormal{-}}6\times10^8\,\rm{M}_{\odot}$ \citep{helmi2018}, $2\mathit{\textnormal{-}}4\times10^8\,\rm{M}_{\odot}$ \citep{kruijssen2020}, $3\times10^8\,\rm{M}_{\odot}\mathit{\textnormal{-}}10^9\,\rm{M}_{\odot}$ \citep{mackereth2019}, $5\mathit{\textnormal{-}}6\times10^8\,\rm{M}_{\odot}$ \citep{alvar2018, vincezo2019}, $2\mathit{\textnormal{-}}5\times10^8\,\rm{M}_{\odot}$ \citep{mackereth2020}, $7\times10^8\,\rm{M}_{\odot}\mathit{\textnormal{-}}7\times10^9\,\rm{M}_{\odot}$ \citep{feuillet2020}, $4\mathit{\textnormal{-}}7\times10^8\,\rm{M}_{\odot}$ \citep{naidu2020}, $6\mathit{\textnormal{-}}8\times10^8\,\rm{M}_{\odot}$ \citep{han2022}, $1\mathit{\textnormal{-}}2\times10^8\,\rm{M}_{\odot}$ \citep{lane2023}, etc.
\citet{das2020} identify a sample of purely accreted stars which they associate with the GSE progenitor. They argue that the kinematic properties of this sample support a larger progenitor mass of $\sim3.4\times10^{11}\,\rm{M}_{\odot}$, from which they infer a stellar mass of $\sim3\times10^9\,\rm{M}_{\odot}$. To summarise, there is general agreement that the progenitor stellar mass of the GSE was between that of the Small and Large Magellanic Clouds, but with a large scatter between different estimates of roughly an order of magnitude. \par

% Talk about the stellar masses:
In comparison to these literature estimates, the stellar masses of the GSE-type mergers in {\sc auriga} are rather high (see the stellar masses in Table \ref{tab:gse_analogue_properties}). \citet{Auriga} compare the stellar to halo mass relations for the central {\sc auriga} galaxies against the semi-empirical relations of \citet{moster2013}. They find that most of the MW-mass galaxies are within the $1\sigma$ scatter of the semi-empirical model at $z=0$, but lie predominantly above the scatter at $z=3$. This was ruled to be a consequence of the sub-grid physics model employed in {\sc auriga}, which lacks sufficient feedback to regulate star formation at earlier times. The discrepancy is even greater for lower-mass galaxies, with dwarfs in the range $10^8<M_{\rm \star}/\text{M}_{\odot}<10^9$ lying above the semi-emprical model by almost $3\sigma$ at $z=0$. \par

As a result, we expect that all simulated galaxies presented here have a greater stellar mass than would be expected from the relations in \citet{moster2013}, and this should be considered carefully before making any direct comparisons to the GSE. However, we note that the mass-to-light fractions remain very high in these galaxies despite their raised stellar masses, and so their dissolution within the host MW-like galaxies is likely to be unaffected. Furthermore, there is already a high scatter in the literature predictions for the GSE progenitor mass, and a scatter in the stellar to halo mass relations from \citet{moster2013} (especially in this mass regime), and so these simulations remain useful for the investigative purposes here. Therefore, we do not expect this to change the qualitative nature of our results, but it will systematically shift properties like the stellar mass and chemical enrichment. \par

\subsubsection{Merger infalls}

Many of the literature works cited in Section \ref{sec:masses} also include an estimate for the original accretion time of the GSE, ranging from $2>z>1$, which translates to lookback times of approximately $10.5>t_{\rm lookback}/\rm{Gyr}>7.9$. This wide range of accretion times may in part reflect the prolonged interaction period between the GSE and the MW (as reasoned in \citealt{naidu2021}). \par

We illustrate the infall of each GSE-type merger in Figure \ref{fig:paper_infall}. The line thickness indicates the stellar mass ratio between the GSE-type merger and the central {\sc auriga} galaxy. The line colour indicates the metallicity difference within twice the stellar half mass radius of each object.

All mergers infall on highly radial trajectories, with orbital circularities (see Equation \ref{equ:e_circ}) of approximately 0. Each merger, except for Au-15-M, falls to within $10\,$kpc on the first pericentre passage. These steep infalls are responsible for the high radial velocity anisotropy of the debris seen in Figure \ref{fig:paper_pie}, and are a characteristic property of GSE-type mergers in general\footnote{Radial GSE-type mergers occur across much of the {\sc auriga} simulation suite, but in other cases there are other major mergers which either overwhelm or directly disturb the radially anisotropic debris footprints.}. The first apocentres range from between 25 and 70\,kpc, whereas breaks in the MW halo and anisotropy profiles suggest that the GSE apocentre occurred at between 25 and 30\,kpc \citep[e.g.][]{deason2018, lancaster2019}. The merger remnants are rapidly disrupted by the host potential, with the stellar mass ratios dropping below $1:100$ within around $2\,$Gyr. Most of the merger progenitors are less iron-enriched and have between four and ten times less stellar mass than the corresponding central {\sc auriga} galaxies. \par

The star formation in each merging galaxy is rapidly quenched upon infall, whereas the central {\sc auriga} galaxies experience an excitation in their star formation rates. Even so, in some cases the metallicity of the merger appears to grow at a faster rate than the central {\sc auriga} galaxy. In fact, this is an illusion caused by the tidal dissolution of the merger: the metal-poor outskirts of the merging galaxy are preferentially stripped, leaving the metal-rich core intact. \par

\subsection{GSE-type merger properties} \label{sec:prog}

We now investigate the diversity in the main GSE-type mergers, in terms of their pre-infall progenitor galaxies and their resulting debris. \par

% Disciness:
%%%%%%%%%%%%%%%%%%%%%%%%%%%%%%%%%%%%%%%%%%%%%%%%%%%%%%%%%%%%%%%%%%%%%%
\subsubsection{$L_z$ distributions} \label{sec:discs}

\begin{figure}
\centering
  \setlength\tabcolsep{2pt}%
    \includegraphics[keepaspectratio, trim={0.0cm 0.0cm 0.0cm 0.0cm}, width=\columnwidth]{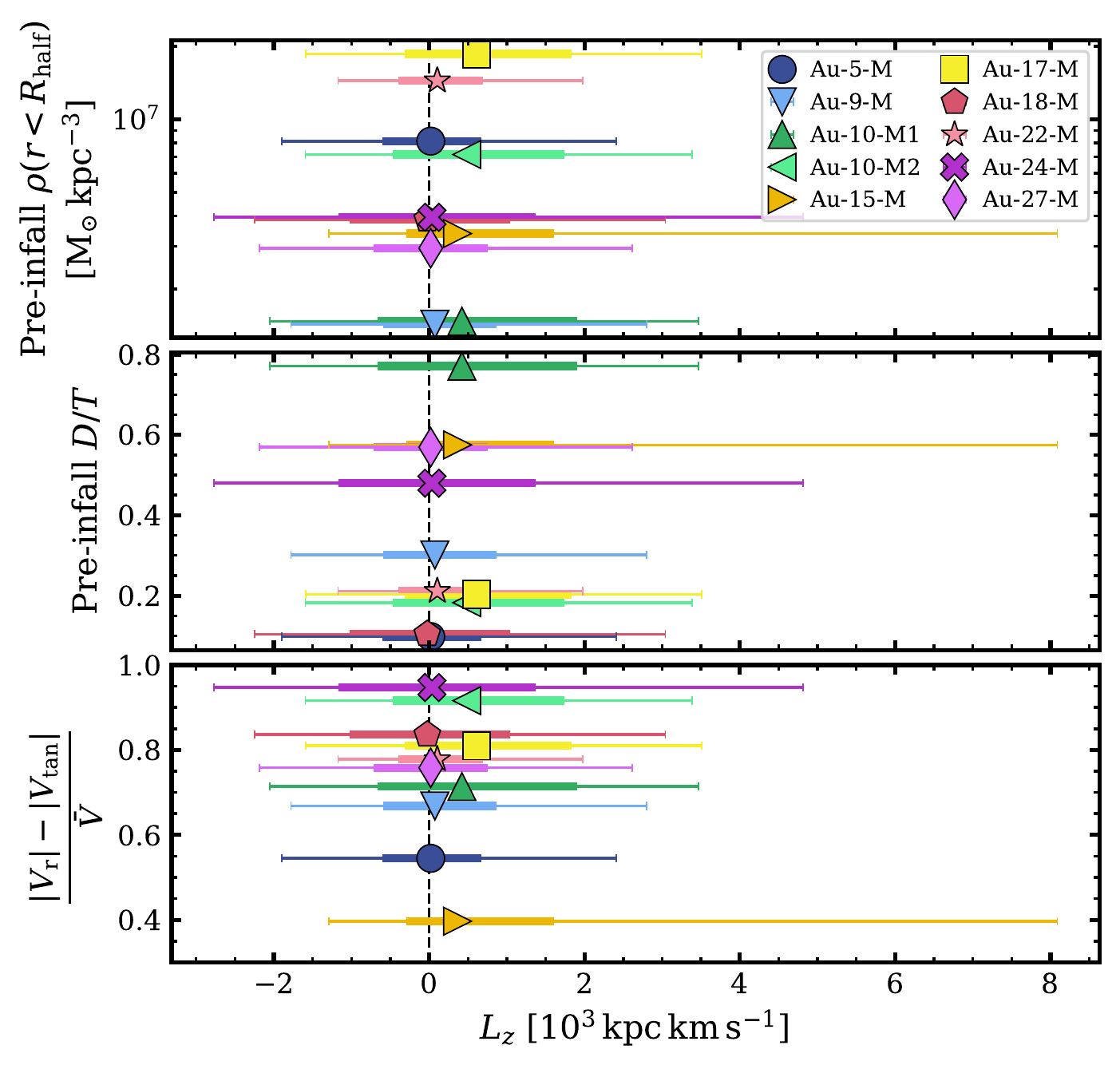}\\
\caption{\textit{Upper panel}: The pre-infall total mass density of each GSE-type merger within the stellar half mass radius, versus the median $z$-component of the angular momentum of the debris at $z=0$.
\textit{Middle panel}: The pre-infall disc to total stellar mass ratio of each GSE-type merger, versus the median $z$-component of the angular momentum of the debris at $z=0$.
\textit{Lower panel}: The normalised radial orbital velocity minus the tangential orbital velocity, recorded at the moment when the merger progenitor crosses the $R_{200}$ radius of the central {\sc auriga} galaxy. A value of $+1$ corresponds to radial orbits, and $-1$ to circular orbits.
The errorbars correspond to the $\pm1\sigma$ and $\pm2\sigma$ percentiles of the distribution, whereas coloured markers correspond to the medians.
There are no overwhelming relationships between pre-infall density or disc fraction and the angular momentum distribution in the debris.}
\label{fig:merger_discs}
\end{figure}

Using controlled idealised simulations of GSE-type merger events, \citet{amarante2022} show that the angular momentum of merger debris is related to the strength of the stellar feedback. Stronger feedback inhibits the recovery of the central density after pericentre passages, hastening the dissolution of the merger and therefore affecting the distribution of its debris in $L_{z}$. This leads to debris with a wider and more asymmetric spread in $L_{z}$, and offset from $L_{z}=0$. Whether the merger is pressure-supported or rotationally-supported may also influence the final distribution \citep[e.g.][]{koppelman2020}. \par

Here, we investigate whether these merger properties have a meaningful influence on the final distribution of their angular momenta. We estimate the disc fraction in each GSE-type merger at the snapshot before $z_{\rm infall}$ following the methodology described in Section \ref{sec:post}, at which time we assume there is minimal tidal disruption to the inner galaxy. The fraction of disc to total stellar mass ($D/T$) is listed for each GSE-type merger in Table \ref{tab:gse_analogue_properties}. We calculate the central density of the GSE-type merger at this same time, evaluated for all matter within the stellar half mass radius. There is a wide range in both of these properties. \par
%Check: https://arxiv.org/pdf/2006.07620.pdf

We show the pre-infall central density (upper panel) and $D/T$ values (middle panel) versus the angular momentum distribution ($L_z$) of the merger debris at $z=0$ in Figure \ref{fig:merger_discs}. The thin and thick errorbars represent the $\pm1\sigma$ and $\pm2\sigma$ percentiles of the $L_{z}$ distribution, thereby indicating both the spread and asymmetry. The coloured markers represent the median of the distribution. There is no clear relationship between the pre-infall central density and the $L_z$ distribution, nor between the pre-infall disc fraction and the $L_z$ distribution. Normalising the results by the mass or density of the central {\sc auriga} galaxy does not affect this outcome. \par

The simulations in \citet{amarante2022} are idealised and non-cosmological, whereas these {\sc auriga} realisations represent a wide diversity in both merger and host properties. Therefore, trends relating $L_z$ to the central density and/or disc fraction of the progenitor galaxies may be overwhelmed by the inherent stochasticity in their other properties. This shows that, at least in a varied cosmological setting, the $L_z$ distribution of merger debris is not highly dependent on the central density or disc fraction. \par

The $L_{z}$ distribution of merger debris can also be affected by the properties of the merger infall. Idealised simulations in \citet{naidu2021} highlight a correlation with the orbital circularity of the merger: increasingly circular orbits shift the debris towards more retrograde (positive) $L_z$. The mergers in our sample are on highly radial infall trajectories, although Au-15-M is an outlier with by far the least radial infall (see Figure \ref{fig:paper_infall}). We evaluate this in terms of the velocity of merger progenitor at infall, and show the result in the lower panel of Figure \ref{fig:merger_discs}. The debris from Au-15-M is notably shifted into the retrograde side of $L_z$, as would be expected from \citet{naidu2021}, but the rest of our sample exhibits no clear relationship. \par

The stellar velocities of the Au-15-M debris is shown in Figure \ref{fig:paper_pie}, which reveals that it is the least radially extended of all the GSE-type mergers shown here. This suggests that Auriga requires mergers with lower orbital circularity than in \citet{naidu2021} in order to produce debris with radial anisotropy comparable to the GSE. \par

% Chemistry
%%%%%%%%%%%%%%%%%%%%%%%%%%%%%%%%%%%%%%%%%%%%%%%%%%%%%%%%%%%%%%%%%%%%%%
\subsubsection{Metallicity gradients} \label{sec:gradient}

\begin{figure*}
\centering
  \setlength\tabcolsep{2pt}%
    \includegraphics[keepaspectratio, trim={0.0cm 0.0cm 0.0cm 0.0cm}, width=0.7\linewidth]{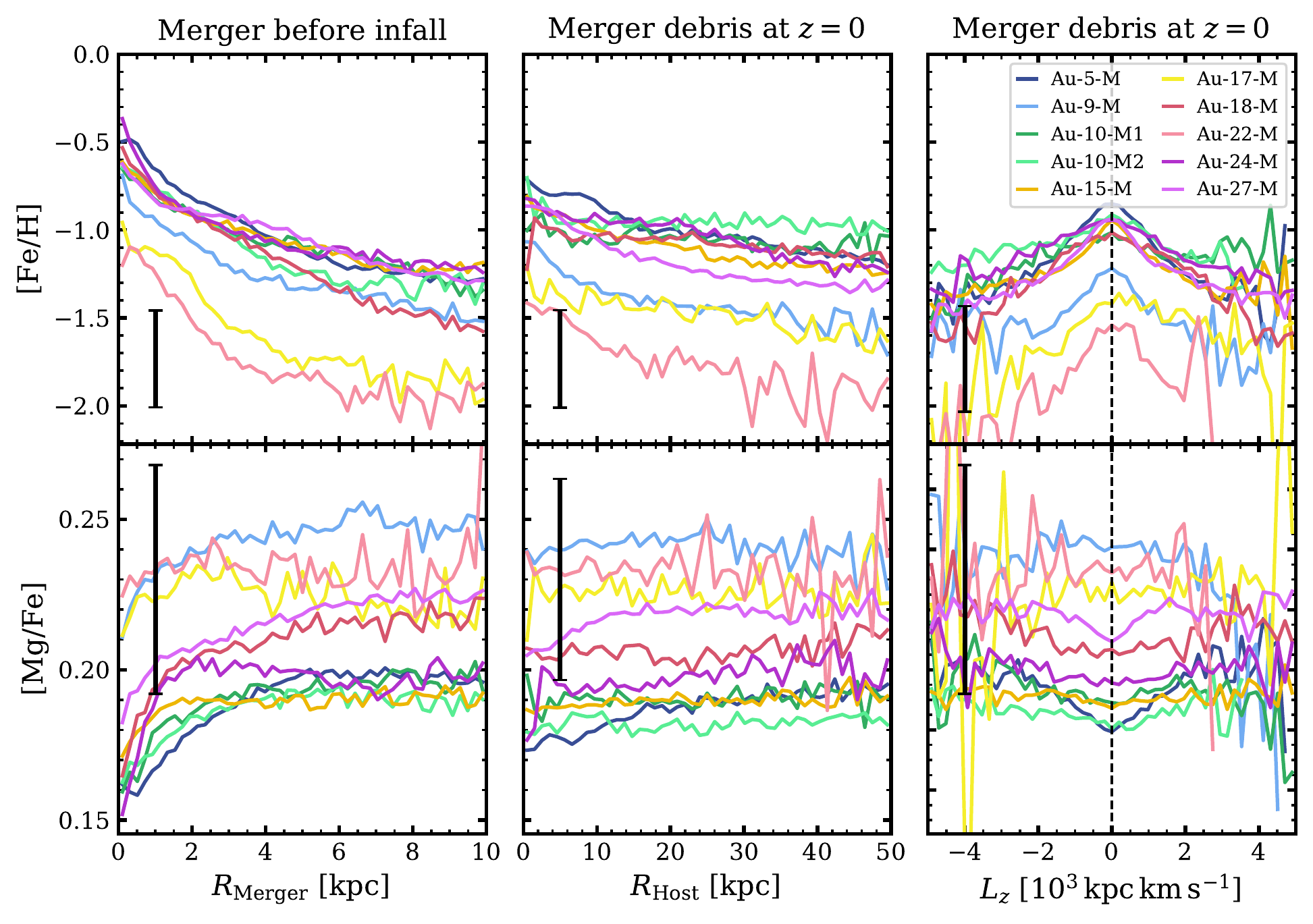}\\
\caption{\textit{Left panels}: The pre-infall median [Fe/H] (upper panel) and [Mg/Fe] (lower panel) radial gradients, shown out to a maximum radius of 10\,kpc. A single black error bar is included to represent the typical standard deviation within 5\,kpc.
\textit{Middle panels}: The radial metallicity gradient of the merger debris at $z=0$, out to a maximum radius of 50\,kpc. Here, the error bar represents the typical standard deviation within 20\,kpc.
\textit{Right panels}: The metallicity gradient of the merger debris but in terms of the $L_z$ angular momentum.
There is not necessarily a consistent mapping between the abundance gradients before infall and in the debris at $z=0$.}
\label{fig:metal_gradient}
\end{figure*}

The hierarchical formation of galaxies generally results in a radial metallicity gradient, with the most evolved stars forming in the galactic centre and less evolved stars accreting at higher radii. As merging galaxies are tidally stripped during infall, the most tightly bound stars are shielded until the merger remnant has decayed to lower orbital energies. Therefore, a pre-infall metallicity gradient should give rise to a corresponding gradient within the merger debris. \par

We show the [Fe/H] (upper panels) and [Mg/Fe] (lower panels) metallicity gradients for each GSE-type merger in Figure \ref{fig:metal_gradient}. The first column shows the radial gradients in the GSE-type merger progenitor at the snapshot before $z_{\rm infall}$. The second column shows the resulting radial gradient of the merger debris at $z=0$. The third column shows the gradient of the merger debris once more, but in terms of the angular momentum $L_z$. In each panel, a single black error bar is used to indicate the typical standard deviation of each stellar metallicity distribution. \par

There is a clear metallicity gradient within the all GSE-type mergers progenitors. The slopes of these gradients are steepest within $\sim5\,$kpc (which is typically comparable to the stellar half mass radius), spanning the range $-0.2<\Delta \text{[Fe/H]}/\Delta R_{\rm Merger}(\rm{kpc})<-0.06$ and $0.002<\Delta \text{[Mg/Fe]}/\Delta R_{\rm Merger}(\rm{kpc})<0.008$. The slopes of the [Fe/H] gradients are only slightly shallower outside of $\sim5\,$kpc, whereas the [Mg/Fe] gradients are flattened in most cases. There is no clear relationship between these slopes and the stellar mass of the merger. \par

The slope of the $z=0$ debris gradients are shallower than before infall, and steepest within the inner $\sim 20\,$kpc, spanning the range $-0.04<\Delta \text{[Fe/H]}/\Delta R_{\rm Host}(\rm{kpc})<0.0$ and $0<\Delta \text{[Mg/Fe]} / \Delta R_{\rm Host}(\rm{kpc})<0.002$. For some of the realisations, the [Mg/Fe] gradients are almost completely flat over all radii. \par

This change in slope is due to three effects: i) the stars are spread over a much wider radial range; ii) the merger is not necessarily dissolved within a single infall, leading to a smearing of the pre-infall gradient over multiple pericentre passages; iii) the distribution of stellar orbits is more radially biased than they were in the pre-infall merger galaxy, meaning some stars that were originally liberated at high-radii are seen at the low-radii pericentres of their new orbits. These effects are not consistent between each GSE-type merger, and the metallicity gradients in some examples are flattened more than others. A similar investigation and discussion is performed for {\sc auriga} galaxies in \citet{monachesi2019}.\par

As highlighted in \citet{naidu2021}, there is evidence for a weak metallicity gradient in the $L_z$ angular momenta of GSE stars. In the third column, we show the metallicity gradients in the $L_z$ angular momentum distributions for each GSE-type merger debris. These gradients are roughly symmetric about $L_z=0$ for $|L_z|<2$, but with some deviations outside of this regime. The (absolute) gradient slopes span the range $0.05<\Delta \text{[Fe/H]}/\Delta L_z(10^3\,\rm{kpc}\,\rm{km}\,\rm{s}^{-1})<0.22$ and $0.0<\Delta \text{[Mg/Fe]} / \Delta L_z(10^3\,\rm{kpc}\,\rm{km}\,\rm{s}^{-1})<0.01$. \par

\subsubsection{The shape of GSE-type merger debris} \label{sec:shape}

\begin{figure*}
\centering
  \setlength\tabcolsep{2pt}%
    \includegraphics[keepaspectratio, trim={0.0cm 0.0cm 0.0cm 0.0cm}, width=\linewidth]{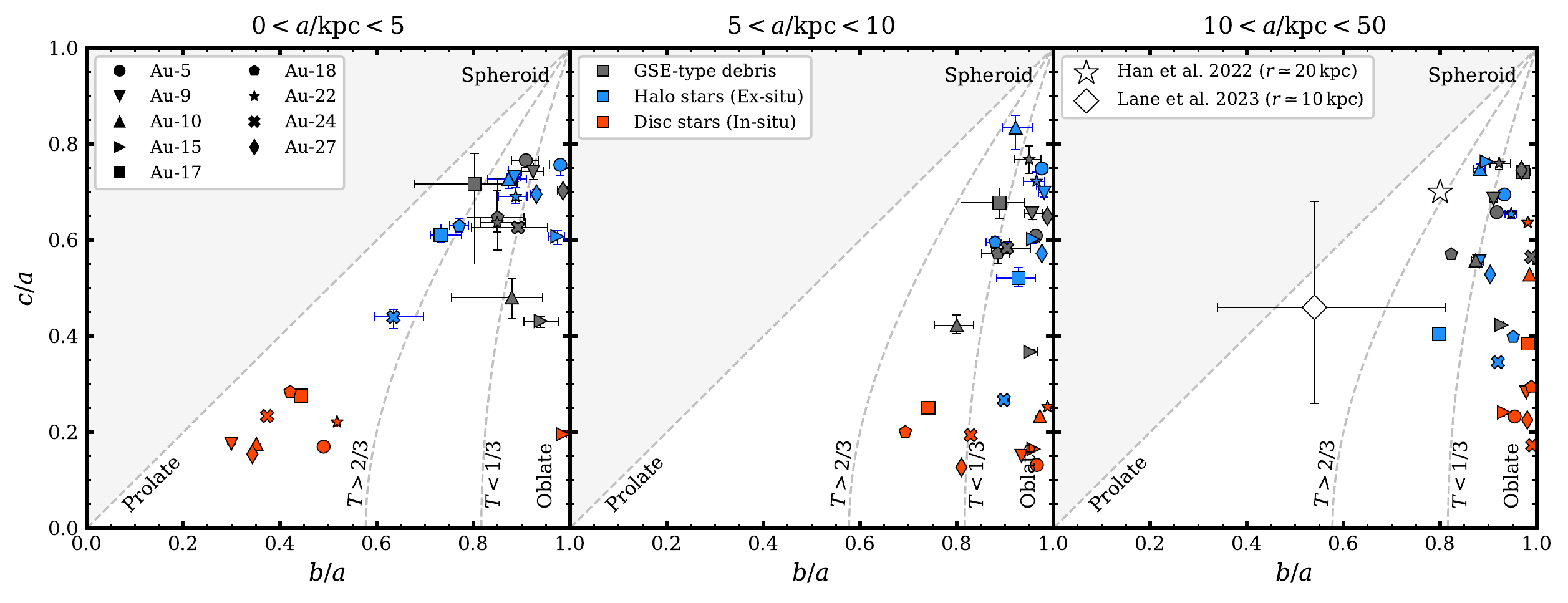}\\
\caption{The shape of different stellar components in each {\sc auriga} galaxy at $z=0$, as measured by the axial ratios $b/a$ and $c/a$. Black markers indicate the shape of stellar debris from GSE-type merger events. For the purposes of this plot, the double-mergers in Au-10 are considered together. Red markers indicate the shape of kinematically selected disc stars. Blue markers indicate the shape of kinematically selected halo stars from an \textit{ex-situ} origin, excluding stars that originated from the GSE-type merger. The shape is calculated in three radial bins, corresponding to the three figure panels, which represent the inner-galaxy ($0<a/\rm{kpc}<5$), the Solar neighbourhood ($5<a/\rm{kpc}<10$), and the outer-galaxy ($10<a/\rm{kpc}<50$). Error bars denote the $\pm1\sigma$ uncertainty derived from a bootstrap method. Grey dashed lines depict the regions that are prolate ($T>2/3$) and oblate ($T<1/3$). The white star in the right panel is based on observed GSE stars from \citet{han2022}, and the white diamond on observed GSE stars from \citet{lane2023}. The shape of GSE-stars is oblate in most cases, except for within the inner 5\,kpc.}
\label{fig:shapes}
\end{figure*}

The ellipsoidal shape of purely collisionless merger debris is known to relate to the infall of the progenitor galaxy \citep{moore2004, cooper2010}, with massive radial mergers creating aspherical and prolate shapes that are aligned with the direction of infall. Therefore, it is possible that the stellar debris from GSE-type mergers has a unique shape that can be distinguished from the rest of the \textit{ex-situ} stellar halo. \par

To consider this possibility, we estimate the shape of stellar debris following the methods described in \citealt{katz1991, dubinski1991, warren1992}. This involves solving the moment of inertia tensor for stars within an initially spherical shell:
\begin{equation}
S_{ij} = \frac{\sum_{k}{m_{k} \mathbf{r}_{k,i} \mathbf{r}_{k,j}}} {\sum_{k}{m_{k}}},
\label{equ:shape}
\end{equation}
where the right-hand-side refers to the elements of $S$ in terms of a stellar particle $k$ with mass $m$ and galactocentric position vector $\mathbf{r}$.
The axial ratios of each volume can be derived from the eigenvalues of $S$, and the orientation of the volume can be derived from its eigenvectors. We iterate this procedure, each time updating the initial ellipsoidal shell with the shape from the previous iteration, until a convergence criterion is met. In this case, we define convergence as when the axial ratios between iterations vary by less than 0.1 per cent. We apply a bootstrap method to this algorithm, where we resample the stellar distribution 100 times using the {\sc resample} function in the {\sc sklearn} {\sc python} package \citep{scikit-learn}. We use the argument ``replace=True'', which replaces a subset of the data array with a random sample of the data. We estimate the best value and $\pm1\sigma$ uncertainties using percentiles. \par

The resulting axial ratios for each {\sc auriga} simulation at $z=0$ are shown in Figure \ref{fig:shapes}. We consider three distinct galactic components: the stellar debris from the main GSE-type merger (black), the remaining \textit{ex-situ} stellar halo (blue) and the \textit{in-situ} disc (red). Disc stars are kinematically selected following the methodology described in Section \ref{sec:post}. The figure is then divided into three panels corresponding to the inner-galaxy ($0<a/\rm{kpc}<5$), the Solar neighbourhood ($5<a/\rm{kpc}<10$), and the outer-galaxy ($10<a/\rm{kpc}<50$), where $a$ is the major axis length. The GSE-type merger debris tails off towards and beyond 50\,kpc, so realistically the outer-galaxy shape calculation will be dominated by stars in the $10\mathit{\textnormal{-}}30$\,kpc range. Grey dashed lines mark constant $T$, where $T$ is the triaxiality parameter defined in \citet{franx1991} as:
\begin{equation}
    T = \frac{1-b^{2}/a^{2}}{1-c^{2}/a^{2}},
    \label{equ:triaxiality}
\end{equation}
where $T>2/3$ is prolate (cigar-shaped) and $T<1/3$ is oblate (pancake-shaped). These lines converge at $c/a=b/a=1$, where the shape is maximally spheroidal. \par

\textit{Inner-galaxy (left panel)}: Most of the disc stars are prolate due to the presence of rotating stellar bars \citep[see][]{blazquez2020, fragkoudi2020}. Au-15 is a solitary outlier, and this is the single {\sc auriga} galaxy from our sample that does not possess a stellar bar. The \textit{ex-situ} stellar haloes are more spheroidal than the disc, and on the whole do not strongly favour prolate or oblate forms. The GSE-like debris has a similar shape to the \textit{ex-situ} halo, though with large uncertainties due to the low number of stellar particles compared to the other components. \par

\textit{Solar neighbourhood (middle panel)}: The disc components have shifted closer to an oblate shape, marking the transition from bar-dominated to disc-dominated. The \textit{ex-situ} haloes and GSE-type merger debris are in most cases highly oblate. \par

\textit{Outer-galaxy (right panel)}: The few remaining disc stars are converged at a highly oblate form. The \textit{ex-situ} halo and GSE-type merger debris have similar overall shapes as they did at the Solar radius, but are less strongly converged at highly oblate configurations. \par

Over all three radial regimes, there is no clear separation between \textit{ex-situ} haloes and GSE-type merger debris across our simulation selection. However, there are some notable differences between the components in \textit{individual} realisations. \par

The \textit{ex-situ} halo in Au-24 has shape approaching that of the disc stars across all three radial regimes. This halo is investigated in detail in \citet{gomez2017b}, where it is shown that a substantial \textit{ex-situ} disc forms from the accretion of mergers on tangentially-biased orbits that are preferentially aligned with the disc plane. The GSE-type merger also infalls along the disc plane, but does so on an extremely radial trajectory. The \textit{ex-situ} accretions are more isotropically distributed in the other realisations, leading to more spheroidal \textit{ex-situ} haloes. \par

The GSE-like debris in Au-15 is more oblate and less spheroidal than in other realisations. This is because the infall trajectory of Au-15-M is the most tangential of all GSE-type mergers (see Figure \ref{fig:paper_infall}), leading to the debris adopting a pancake shape. \par

The GSE-like debris in Au-10 is less spheroidal than the \textit{ex-situ} halo ($\Delta(c/a)>0.2$), and also less spheroidal than the GSE-type mergers from most other simulations. This is unexpected because the majority of the GSE-like debris in Au-10 is comprised almost equally of stars from two separate merger events, and the superposition of two shapes should favour spheroidal symmetry. However, the two GSE-type mergers infall along nearly polar opposite directions, and their combined angular momenta are complementary. \par

\begin{figure}
\centering
  \setlength\tabcolsep{2pt}%
    \includegraphics[keepaspectratio, trim={0.0cm 0.0cm 0.0cm 0.0cm}, clip, width=\columnwidth]{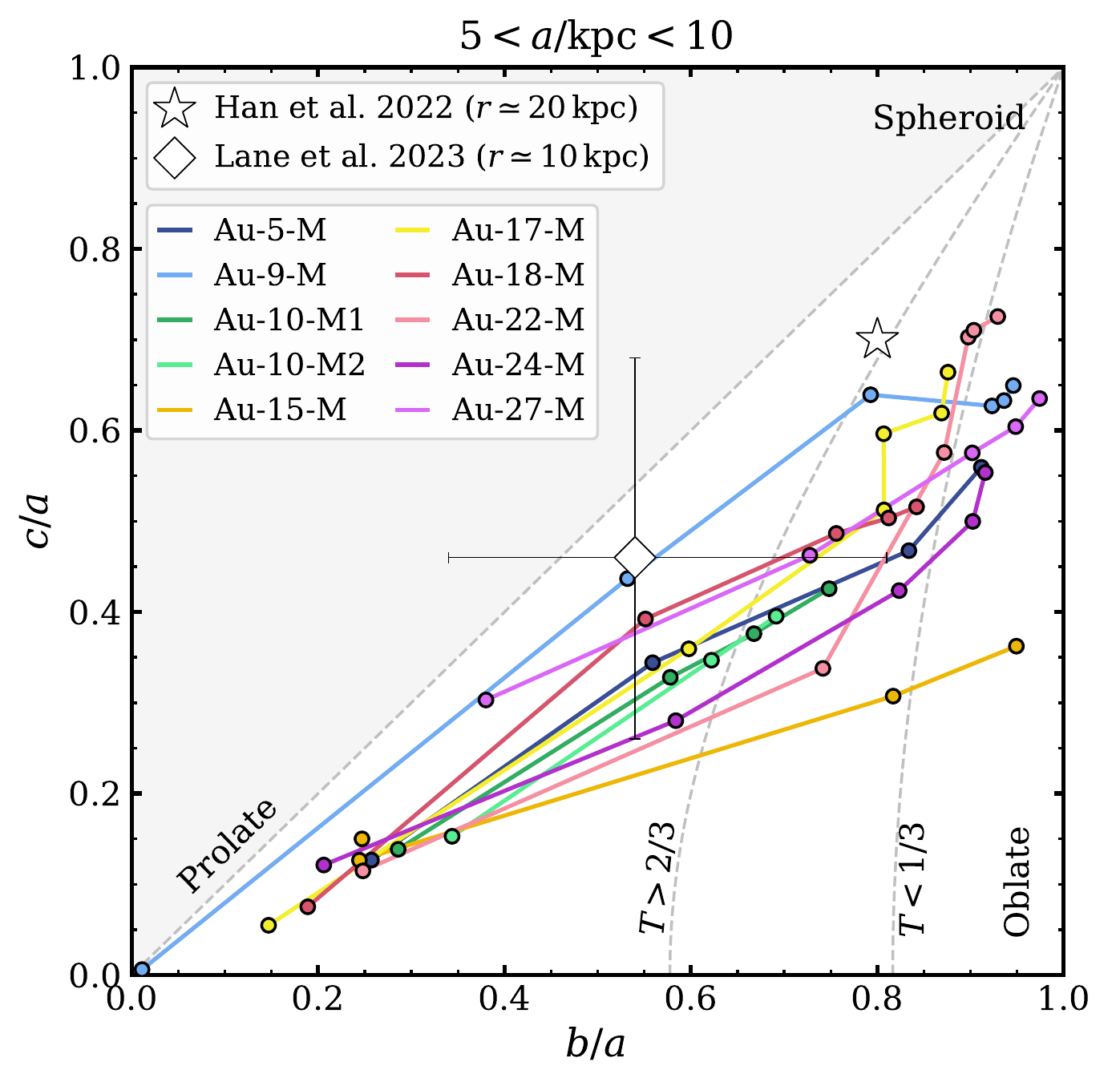}\\
\caption{The shape evolution of the GSE-type stellar debris in the range $5<a/\rm{kpc}<10$, as measured by the axial ratios $b/a$ and $c/a$. These axial ratios are averaged over bins in lookback time with a $2\,$Gyr width which progress from left to right, where each interval is marked by a black circle. The distribution is initially highly prolate in all cases.}
\label{fig:shape_evolution}
\end{figure}

We show the time evolution of the GSE-type debris shape for the `Solar neighbourhood' regime in Figure \ref{fig:shape_evolution}. This demonstrates that the debris from every GSE-type merger event is initially elongated and prolate, exactly as would be expected given their highly radial infall trajectories. Time intervals of $2\,$Gyr are marked with black rings, which shows that the shape progresses towards rounder and increasingly oblate shapes. Whilst not shown here, the evolution of the DM debris shape follows very similar paths. There are several mechanisms that could be contributing to this shape change:
\begin{itemize}
\setlength{\itemindent}{0.5em}
    \item Torques from the galactic disc/bar, as well as the underlying DM halo, will encourage the debris to both align with and adopt the shape of the disc/bar and DM halo. The growth of the stellar disc/bar can itself impact the shape and alignment of surrounding debris \citep[e.g.][]{berentzen2006, debuhr2012}.
    \item The transition between centrally DM dominated and baryon dominated can turn prolate shapes into rounder and more oblate shapes \citep{tomassetti2016}.
    \item A dynamic gravitational potential can perturb the orbits of particles, which transforms prolate shapes into rounder and more oblate shapes. Such shape transformations are widely reported in the DM haloes of galaxies \citep{tissera1998, kaz2010, zhu2017}, including in the {\sc auriga} simulation suite \citep{prada2019}, and the stellar halo will be similarly affected. This process is most efficient in regions of high gas density, although the halo shapes in the ``inner-galaxy'' are not notably more spheroidal than elsewhere.
\end{itemize}
\par
The GSE-type merger debris appears to be influenced by these background galactic processes. In all cases, the stellar disc undergoes rapid growth during the time of the GSE-type merger accretion. The early discs in the central {\sc auriga} galaxies are over-massive because the {\sc auriga} galaxies lies above the typical stellar mass halo mass relation at early times (as highlighted in Section \ref{sec:masses}). As such, the torques from the disc component may have been stronger than in the MW. \par

In the ``outer-galaxy'' panel, we include the axial ratios as derived for GSE stars in \citet{han2022} as a white star. These stars are selected with chemical cuts and a cut on orbital eccentricity of $>0.7$, and the shape is evaluated at a `flattened' radius of 20\,kpc. They find a spheroidal shape which tends towards a prolate configuration. A similar shape is also determined in \citet{iorio2019}. \par

We also include the axial ratios as derived for a high-purity sample of GSE stars in \citet{lane2023} as a white diamond. These are selected with chemical cuts and a selection in action space, and the sample is centered around radii of $\sim10\,$kpc. They find a more elongated shape with axial ratios of $1:0.55:0.45$, but with greater uncertainties due to the smaller size of their sample. \par

It is curious that none of the GSE-type debris in our {\sc auriga} selection are consistent with any of these measurements, though we emphasise that the absolute differences in the axial ratios are small when compared to \citet{han2022}. To investigate this further, we reproduce the shape fits on the GSE-type debris using the same selection criteria as in \citet{han2022}. The resulting shapes at $0<a/\rm{kpc}<5$ and $5<a/\rm{kpc}<10$ are more spheroidal, with raised $c/a$ ratios (by $\Delta(c/a)\simeq0.2$ in the most extreme case) but similar $b/a$ ratios. The difference in the $c/a$ ratios can be attributed to the eccentricity cut, which favourably removes merger stars that have become captured by the disc. However, the shape of the debris at $10<a/\rm{kpc}<50$ is insensitive to the selection criteria. Overall, the resulting shape fits remain inconsistent. \par

\subsubsection{The alignment of GSE-type debris with the \textit{in-situ} disc} \label{sec:alignment}

\begin{figure}
\centering
  \setlength\tabcolsep{2pt}%
    \includegraphics[keepaspectratio, trim={0.0cm 0.0cm 0.0cm 1.0cm}, clip, width=\columnwidth]{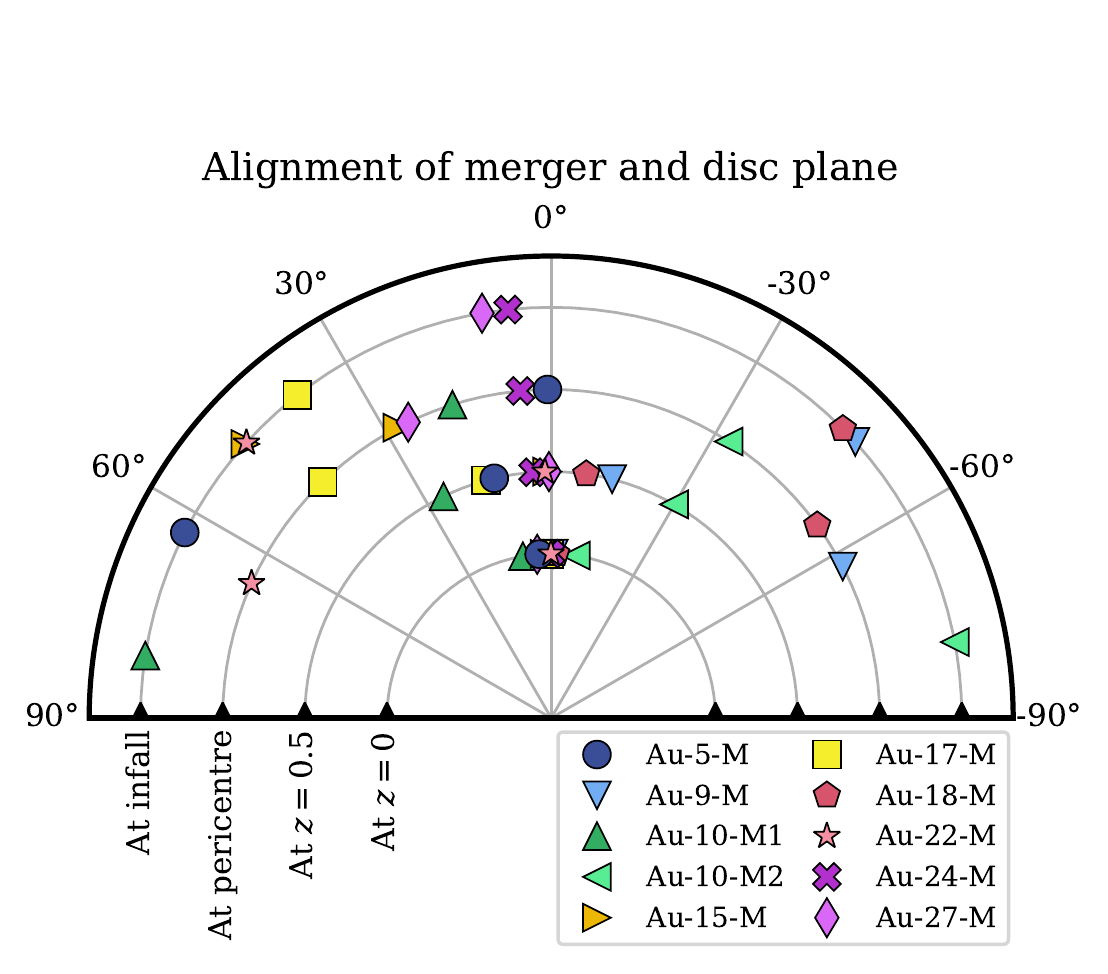}\\
\caption{The alignment angle between the angular momentum vector of the \textit{in-situ} disc and the GSE-type merger, defined such that an angle of 0 degrees corresponds to an exact alignment. We show this alignment at four different times, as indicated by the axis labels. At infall, the alignment of the GSE-type merger is defined by its orbital velocity vector. Otherwise, the alignment is based on the minor-axis of the merger debris, derived using a shape fit to all debris over the radial range $5<a/\rm{kpc}<50$. Regardless of the alignment at earlier times, all GSE-type debris has become aligned with the disc by $z=0$.}
\label{fig:alignment}
\end{figure}

% Introduce topic of alignment.
In addition to estimating the shape of the observed GSE debris, \citet{iorio2019, han2022, lane2023} also estimate the angular tilt with respect to the Galactic disc plane. They find angles of $20$, $25$ and $16$ degrees respectively. \par

% What causes the shape alignment?
One possibility is that this tilt reflects the infall trajectory of the GSE, with the major axis pointing in the direction that the merger originated. In support of this scenario, \citet{chandra2022} detect overdensities in the outer $60\mathit{\textnormal{-}}90\,$kpc of the MW halo, which they link to apocentric shells created during a retrograde and high-inclination GSE merger event. The Virgo Overdensity \citep{vivas2001} and Hercules-Aquila Cloud \citep{belokurov2007b} may also be formed by stars shed during the GSE infall. Conversely, $N$-body simulations in \citet{naidu2021} suggest that the infall trajectory is not \textit{necessarily} preserved in the orientation of the merger debris, finding that an infall inclination of 15 degrees above the disc yields a debris distribution that is inclined at 35 degrees. \par

% In Auriga, the stellar debris and disc plane become aligned over 2 Gyr timescales:
\citet{gomez2017b} investigated the discs in the {\sc auriga} simulations, and showed that the trajectory of massive mergers which contribute to co-rotating stars in the disc plane become aligned with the disc over timescales of $\sim2\,$Gyrs. This is due both to the merger aligning with the disc, and to the disc aligning with the merger, caused by the transfer of angular momentum and dynamical friction \citep[i.e.][]{huang1997, read2008, earp2019}. \par

% Emphasise that we specifically observe this same effect for the GSE-type mergers shown in this work:
We observe a similar effect for the shapes of GSE-type mergers here, continuing long after the merger is disrupted. In Figure \ref{fig:alignment}, we show the angular alignment between the \textit{in-situ} disc and the GSE-type merger, defined such that an angle of $0$ degrees corresponds to alignment and $\pm90$ to perpendicular misalignment. By $z=0$, all GSE-type stellar debris is aligned with the disc to within $\pm15$ degrees. The evolution towards alignment progresses from the inside-out, with debris at low radii aligning most quickly. At higher radii, beyond the influence of the disc, the debris preferentially aligns with the underlying DM halo instead. Over time, this underlying DM halo is itself brought into alignment with the disc. \par

% Why would the debris evolve to be aligned with the disc?
This evolution towards closer alignment could be due to the response of the merger debris to the growth of the stellar disc, an effect which is known to be the most efficient for stars on radial orbits such as these (\citealt{binney1986}, and see also \citealt{dubinski1995}). Additionally, the gas donated by the GSE-type mergers may encourage a tilting of the disc into closer alignment with the merger \citep[as in][]{debattista2015, earp2019}. Similar effects are also seen in the {\sc Artemis} cosmological simulations (\citealp{dillamore2022}, and see also \citealp{dodge2023}). \par

% Further discussion:
Of the GSE-type mergers presented here, Au-10-M1 and Au-10-M2 infall at an almost perpendicular angle to the disc. Nevertheless, by $z=0$ the disc and the GSE-type debris are closely aligned. These results are not consistent with the larger tilt angles found for the GSE in \citet{iorio2019, han2022}, but may still indicate that the modern GSE-disc alignment is not necessarily a reliable reflection of the alignment at infall, or even of the original infall trajectory with respect to cosmic filaments. Whilst \citet{naidu2021} also find a large tilting angle for their best-fit model, their simulations were pure $N$-body and did not model the growth of the stellar disc. If baryonic physics were included, then their merger debris may have become more closely aligned with the disc. \par

% Ramifications:
If torques between the disc and merger debris were operating at this same level in the MW, it may suggest that the GSE accreted later than expected and has not yet become fully aligned with the disc. This is, however, inconsistent with the ancient accretion times inferred by direct measurements of the GSE age \citep[e.g.][]{belokurov2020, bonaca2020}. As already stated, the alignment timescales in {\sc auriga} may be artificially fast due to the high stellar mass at earlier times, which would help to mitigate this problem. \par

\subsubsection{Consideration of the DM halo} \label{sec:DM}

\citet{han2022b} show that both the shape and tilt of stellar debris can be preserved for many Gyrs when immersed in a DM distribution that shares the same shape and tilt. This may be a natural outcome if the merger contributes a meaningful fraction of the local DM. In \citet{iorio2019}, it is estimated that the GSE could contribute as much as 50 per cent of the total DM budget within $30$\,kpc. Then again, other works instead favour a minor-merger scenario in which the GSE would contribute a far lower fraction \citep[e.g.][]{lane2023}. \par

% New text:
The shape and alignments of the DM haloes in {\sc auriga} have previously been investigated in \citet{prada2019} and \citet{gomez2017}. \citet{prada2019} finds typically oblate DM haloes that favour alignment with the stellar disc. In particular, \citet{gomez2017} shows that this alignment is closest for the DM within 10\,kpc, whereas DM at higher radii is increasingly tilted and time-variable in some realisations. This tilting is in most cases due to interactions from massive merger encounters in the last few Gyrs. \par

We also investigate the shape and alignment of DM that originated from the GSE-type mergers. Whilst initially prolate, the shape gradually grows increasingly oblate with time, similar to what is seen for the stellar debris. Interestingly, the alignment of this DM is rapidly coupled to the alignment of the DM in the host galaxy, which in turn is usually aligned with the stellar disc. \par

To summarise, both the stellar debris and DM halo have grown into alignment with the disc by $z=0$, but this timescale may be unrealistically fast due to the high disc mass at early times. The tilting reflects the coupling of the merger debris with the orientation of the host halo, and may not reflect the infall trajectory of the merger itself. \par

\subsubsection{Distinguishing the debris from two progenitors} \label{multimerger}

\begin{figure}
\centering
  \setlength\tabcolsep{2pt}%
    \includegraphics[keepaspectratio, trim={0.0cm 0.0cm 0.0cm 0.0cm}, width=\columnwidth]{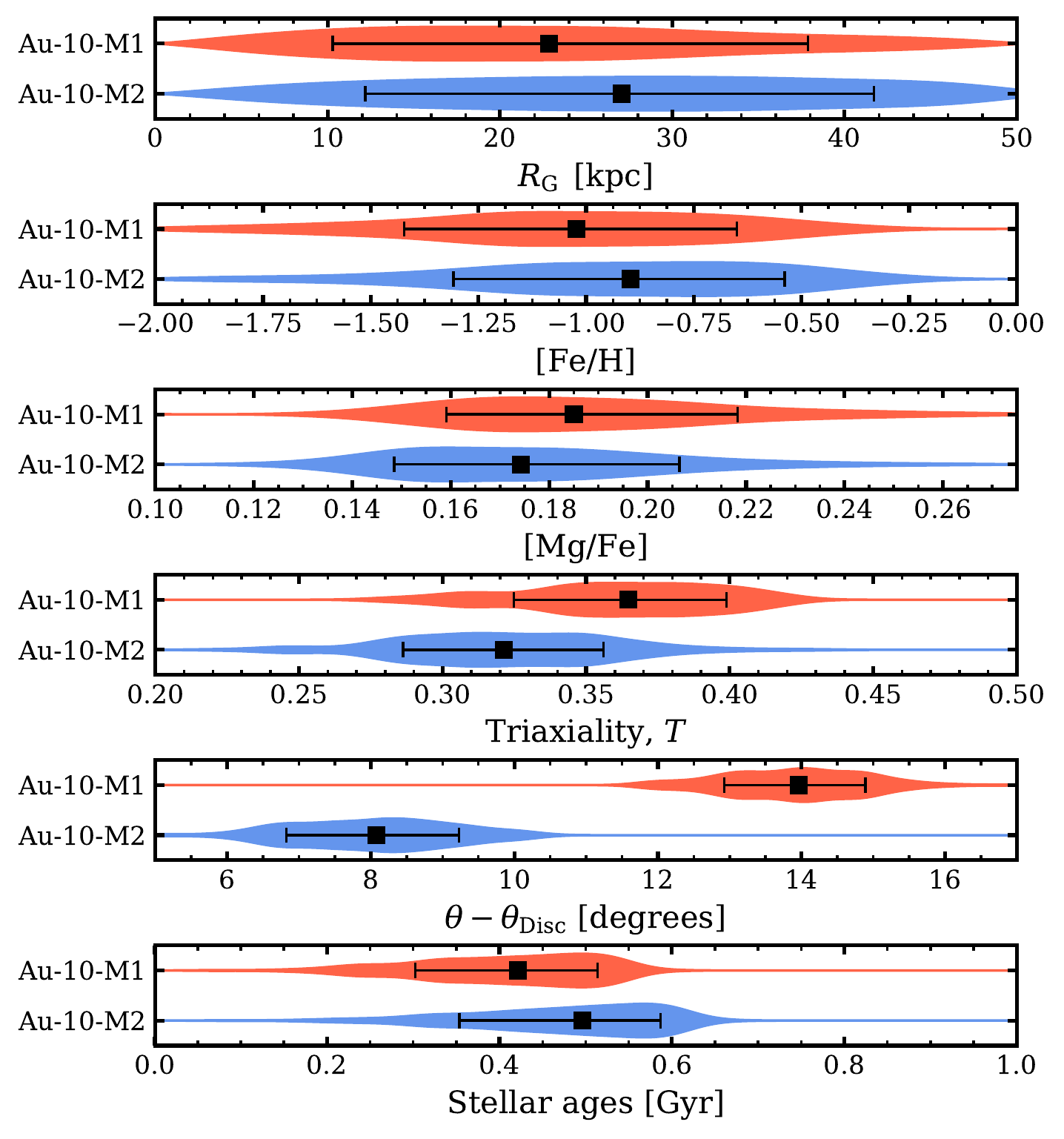}\\
\caption{Violin plots for various properties of the GSE-type debris in Au-10. Black squares indicate the medians of each distribution, and errorbars the $\pm1\sigma$ standard deviation. The coloured bands represent a normalised Gaussian kernel density estimate fit to the simulation data. There are substantial overlaps in the distributions of most properties, which owe to the similar properties of each progenitor galaxy.}
\label{fig:double_merger_comparison}
\end{figure}

As shown in Section \ref{sec:contributions}, the radially anisotropic debris features in {\sc auriga} contain the debris from several merger events, but typically have one dominant originator. Au-10 experiences two mergers that contribute an almost equivalent mass fraction of radially anisotropic stellar debris. Whilst their debris forms similar ellipsoids in $v_{\phi}/v_{r}$ coordinates, they may have distinct distributions in their other properties. Here, we investigate whether it is possible to distinguish their debris at $z=0$. \par

We compare a selection of properties for Au-10-M1 and Au-10-M2 in Figure \ref{fig:double_merger_comparison}. These include the galacto-centric radius of the debris, the [Fe/H] and [Mg/Fe] metallicity distributions, the triaxiality and alignment of the shape with respect to the \textit{in-situ} stellar disc, and the formation times of member stars. We base these comparisons upon the stars present within each merger prior to infall, and over the final radial range $0<R_{\rm G}/\rm{kpc}<50$. For each property, we calculate the percentiles of the distribution and display the results in the form of a violin plot. The median of each property is included as a black square, with errorbars representing the $\pm1\sigma$ percentiles. \par

The medians of the radial, chemical, and shape distributions all overlap within their $\pm1\sigma$ percentile limits. The alignment of the debris shape appears to be distinct due to the narrow percentile distributions, but the absolute difference of $\sim6$ degrees would be exceedingly difficult to detect in practice. The stellar ages are the most effective in differentiating each merger. However, it would remain difficult to discern this difference when considering the superposition of both distributions, especially once observational uncertainties of $\mathcal{O}$(Gyr)  are taken into account \citep[see][]{Miglio2017, carme2019, miglio2021}. These results are comparable to \citet{rey2023}, where it is shown that the debris from several distinct merger events can be difficult to parse. \par

If the merger progenitors are massive and gas-rich, then the gas shocks resulting from their accretion could trigger bursts of \textit{in-situ} star formation \citep[see][]{mihos1994, tissera2002, carme2019, gargiulo2019, ruiz2020b, massana2022, orkney2022}. Indeed, there are excitations in the \textit{in-situ} star formation rate of up to half an order of magnitude during the first pericentre passage of all GSE-type mergers presented in this work. In some cases, there are further excitations resulting from subsequent pericentre passages, but the duration and magnitude of these bursts diminishes rapidly. A multiple-peaked \textit{in-situ} star formation history could be a signature of multiple merger events, but would not necessarily mean that those merger events were GSE-like \citep[e.g.][]{orkney2022}. Furthermore, it would remain difficult to distinguish these peaks if the mergers were accreted within a short time of each other, as is the case here. \par

\subsection{Satellites of GSE-type mergers} \label{sec:sats}

% Satellite mass function
%%%%%%%%%%%%%%%%%%%%%%%%%%%%%%%%%%%%%%%%%%%%%%%%%%%%%%%%%%%%%%%%%%%%%%
\subsubsection{Satellite mass function}

\begin{figure*}
\centering
  \setlength\tabcolsep{2pt}%
    \includegraphics[keepaspectratio, trim={0.0cm 0.0cm 0.0cm 0.0cm}, width=\linewidth]{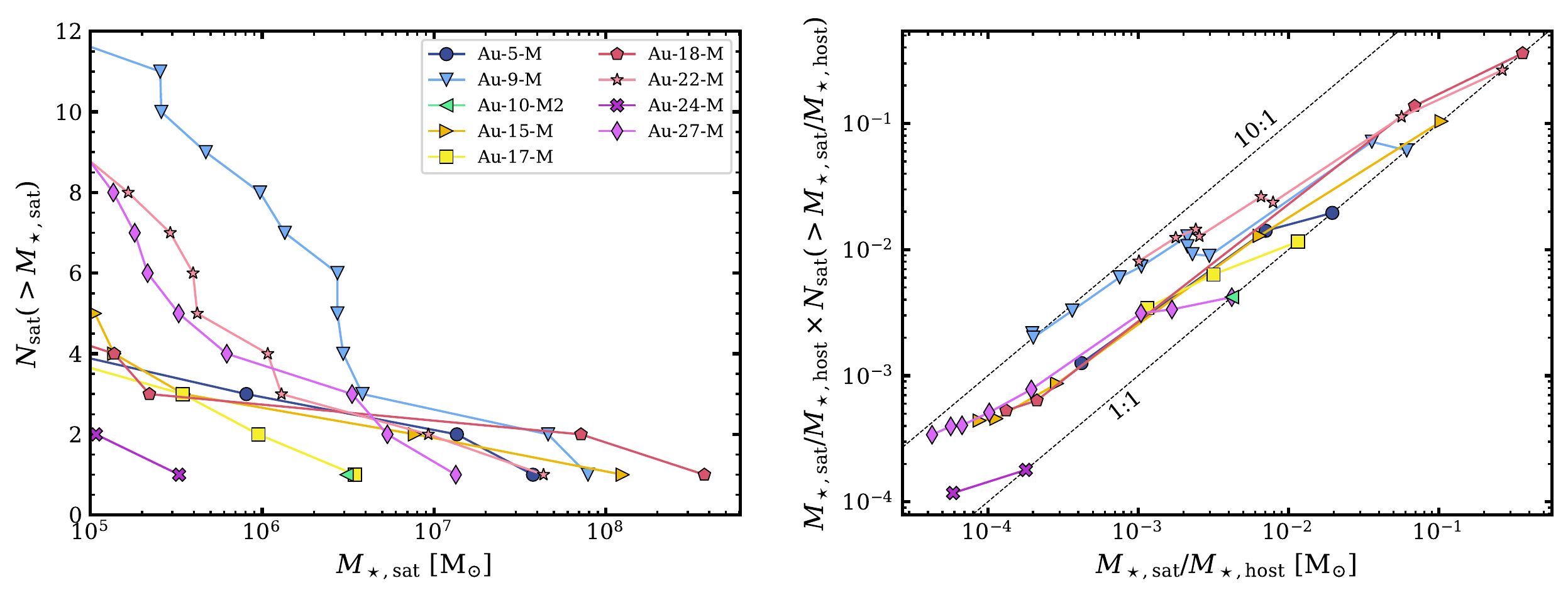}\\
\caption{\textit{Left panel:} The satellite mass distribution function for each of the GSE-type mergers (not including the GSE-type merger itself), down to a lower stellar mass limit of $10^5\,\rm{M}_{\odot}$. The satellite mass is given by the bound stellar mass identified by {\sc subfind}. \textit{Right panel:} The same distribution function, but normalised with respect to the pre-infall stellar mass of the central galaxy in the GSE-type merger system. If the distribution functions were proportional to the stellar mass of the host, then these normalised distributions would overlap. This shows that there is a wide diversity in the satellite mass distribution functions, and at least some of that diversity is independent of the host mass.}
\label{fig:merger_MDF}
\end{figure*}

Investigations of the LMC have suggested that it brought several smaller galaxy companions into the MW \citep[e.g.][]{jethwa2016, sales2017}. Assuming that the GSE progenitor had a stellar mass similar to the LMC as suggested by \citet{das2020} and \citet{evans2022}, then it too may have been accompanied by a number of luminous satellites. If the debris or remnants of these satellites can be identified, then they may help to constrain the properties of the GSE itself. Alternatively, the debris may be mistaken for entirely independent accretion events. \par

Many of the GSE-type mergers presented here host a population of luminous satellites. We identify this population as satellites within the Friends-of-Friends (FoF) group of the pre-infall merger galaxy before it itself falls into the central {\sc auriga} galaxy. Then, we exclude any satellites for which their orbital kinetic energy exceeds the gravitational potential energy binding them to the merger system, and perform a visual check that the satellites remain associated with the merger system until $z_{\rm infall}$. Many of the remaining bound satellites are accreted only a few 100\,Myrs before $z_{\rm infall}$, though there is a selection bias because some satellites that accrete earlier have already dissolved before this time. \par

We show the satellite stellar mass distribution function for each GSE-type merger in the left panel of Figure \ref{fig:merger_MDF}. There is a wide variation in the total number and masses of these satellites, with $0\mathit{\textnormal{-}}8$ satellites at $M_\star>10^6\,\rm{M}_{\odot}$. This may simply reflect the wide variation in the mass of the host GSE-type mergers themselves (see Table \ref{tab:gse_analogue_properties}). We account for this in the right panel, where we normalise the distribution functions with respect to the coeval mass of the GSE-type merger. If the original distribution functions are entirely proportional to the stellar mass of the host, then these normalised distribution functions would overlap. Instead, a moderate degree of scatter remains, with variation in the normalised satellite mass of $\approx4$ orders of magnitude, and variation in the normalised number of $\approx0.5$ orders of magnitude. See \citet{sales2013} for an investigation into the observed satellite stellar mass distribution functions, which finds it to be relatively independent of the host mass for hosts of $M_{\star}<10^{10}\,\text{M}_{\odot}$. \par

% Satellite debris coloured histograms
%%%%%%%%%%%%%%%%%%%%%%%%%%%%%%%%%%%%%%%%%%%%%%%%%%%%%%%%%%%%%%%%%%%%%%
\subsubsection{Satellite debris}

\begin{figure}
\centering
  \setlength\tabcolsep{2pt}%
    \includegraphics[keepaspectratio, trim={0.0cm 0.0cm 0.0cm 0.0cm}, width=\columnwidth]{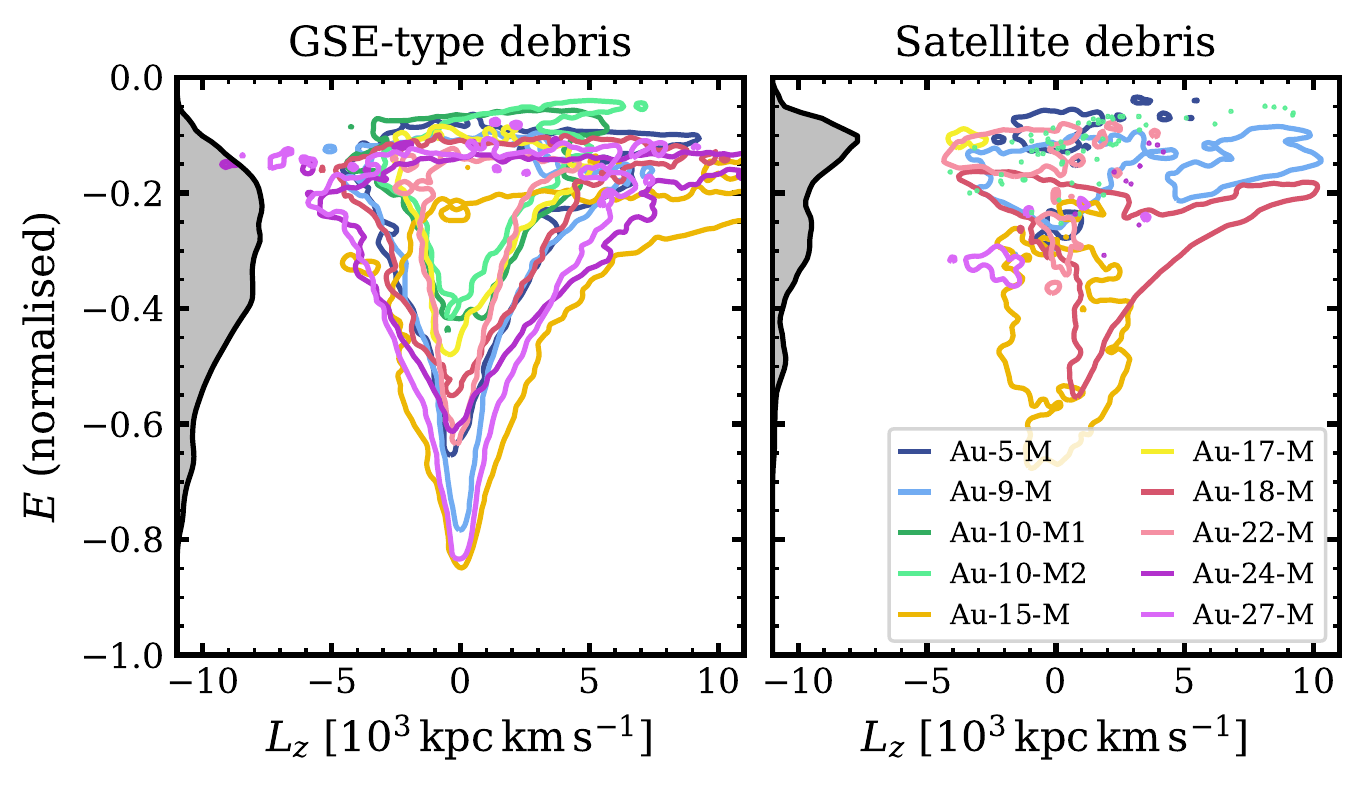}\\
\caption{\textit{Left panel:} Outer contours of the total specific orbital energy versus the specific angular momentum for stars that came from the main galaxy in each GSE-type merger system. \textit{Right panel:} The same, but for stars that came from satellite galaxies of the GSE-type merger system. A grey histogram is included on the $y$-axis which shows the summed PDF across all realisations. Satellite debris is favourably deposited at higher energies, and with a less symmetric distribution in $L_z$.}
\label{fig:sat_debris}
\end{figure}

We have established that GSE-type mergers in {\sc auriga} can host a variable number of luminous satellites, and now we show where those stars are deposited in the central {\sc auriga} galaxies at $z=0$. \par

We show the total specific orbital energy versus the specific angular momentum distributions for each GSE-type merger in the left panel of Figure \ref{fig:sat_debris}. The energy has been normalised in the range $0>E>-1$ to aid comparison, where $E=0$ corresponds to the $R_{200}$ radius, and $E=-1$ is the potential of the host MW-type galaxy at an arbitrarily chosen inner radius ($R_{\rm G}=30\,$pc). The energy distributions of stars that belonged to satellites of the GSE-type mergers are shown in the right panel. We discard any satellite stars that coalesced with the main GSE-type merger prior to it achieving peak stellar mass, and any stars within surviving satellite remnants. \par

The debris from each GSE-type merger spans a wide range of energies that extends from near the virial radius to the Solar radius and below. The debris from some GSE-type mergers reaches far lower energies, and this can be understood from their merger mass ratios in Figure \ref{fig:paper_infall}. The remnants of higher-mass ratio mergers are shielded against tidal disruption for a longer time, and also infall more rapidly due to a greater dynamical friction force\footnote{The dynamical friction timescale is proportional to the merger mass ratio \citep{binney1987}.}. \par

In most cases, the satellite debris is preferentially deposited at higher average energies. This is because the satellites are among the first stars to be stripped from the merging system. Once the satellites are independent of the GSE-type system, the dynamical friction they experience is greatly reduced and they start dissolving due to gravitational tides. This leads to the deposition of their debris at similar energies to where they were originally stripped. Despite this, the absolute density of the satellite debris is often overwhelmed by the GSE-type merger debris at the same energies. \par

\begin{figure*}
\centering
  \setlength\tabcolsep{2pt}%
    \includegraphics[keepaspectratio, trim={0.0cm 0.0cm 0.0cm 0.0cm}, width=\linewidth]{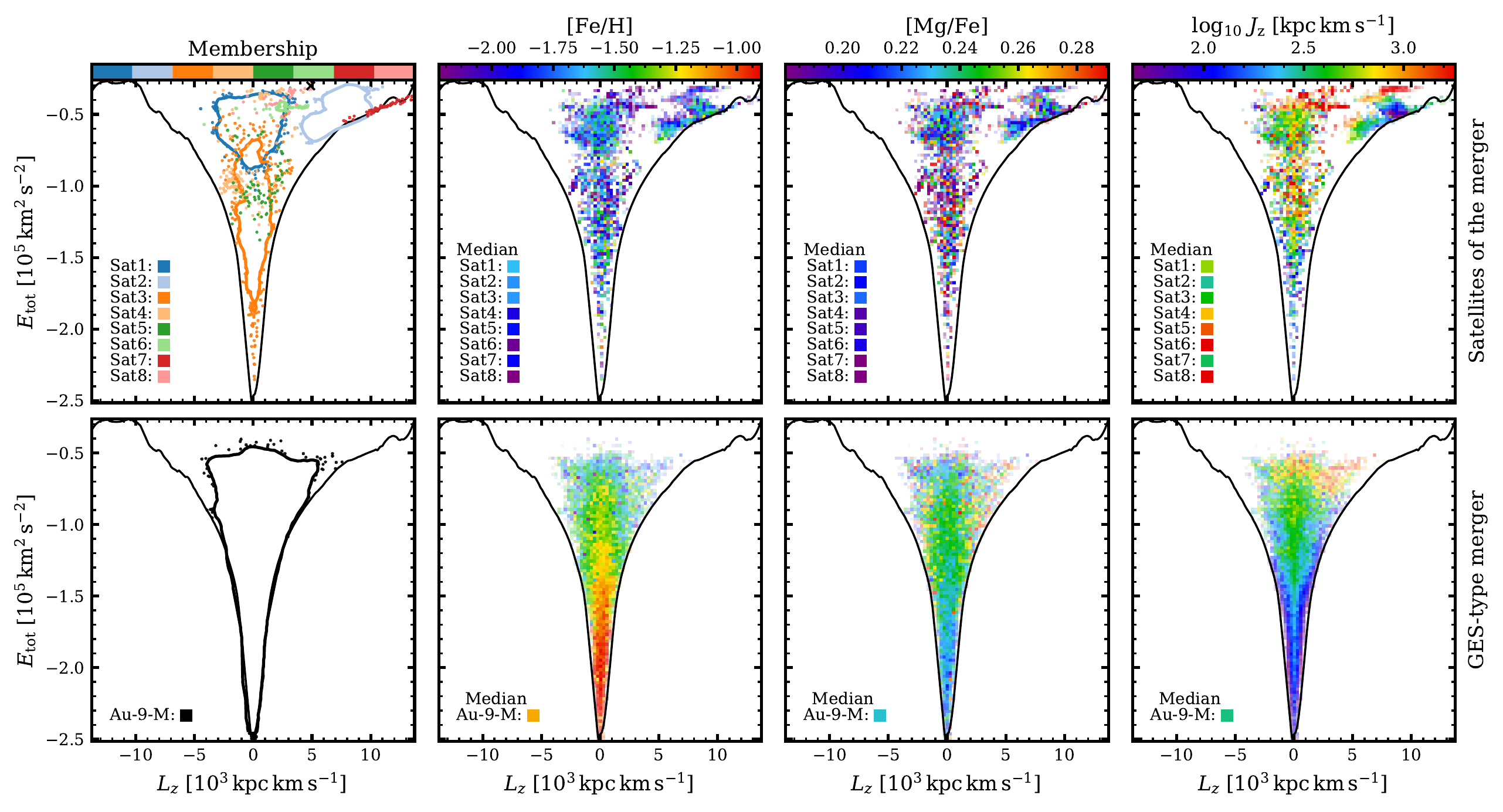}\\
\caption{The total specific orbital energy versus the $z$-component of the specific angular momentum for the Au-9-M GSE-type merger. A black line marks the smoothed edge of the total stellar distribution in Au-9. \textit{Upper panels:} Stars that were exclusively associated with $M_{\star}>10^6\,\rm{M}_{\odot}$ satellite galaxies of the GSE-type merger prior to infall. \textit{Lower panels:} Stars that were associated with the main GSE-type merger at the time of its peak stellar mass. The left-most column groups the stars by their pre-infall galaxy (contours and points). The presence of a surviving satellite remnant (Sat8) is indicated with a black cross. The following three columns include stacked, mass-weighted histograms coloured by the metallicity ([Fe/H), the $\alpha$-metallicity ([Mg/Fe]), and the $z$-action ($J_{\rm z}$). The histogram opacity is based on the mass within each bin. The legends list the median property value for each distinct group. Whilst there is a large gap between the median properties of the satellites and the main merger, these differences are minimised in the energy-space regions where the populations overlap.}
\label{fig:sat_hists}
\end{figure*}

Next, we perform a more detailed investigation into one example {\sc auriga} simulation. We choose to focus on Au-9-M, because this object has the greatest number of satellite galaxies over a wide range of stellar masses. We consider only the satellites that had a stellar mass $>10^6\,\rm{M}_{\odot}$ prior to infall, yielding eight objects. We show merger debris in the plane of the total specific orbital energy against the $z$-component of specific angular momentum in Figure \ref{fig:sat_hists}. The upper panels show stars that were within satellites of Au-9-M prior to its infall, excluding any which became bound to Au-9-M at the time of its peak stellar mass (but not any which became bound after this time). The lower panels show all stars that were within the GSE-type merger at the time of its peak stellar mass. \par

The median chemical abundances of Au-9-M and each of its satellites span approximately 1 dex in [Fe/H] and 0.05 dex in [Mg/Fe], in line with expectations from their differing stellar masses and evolutionary stages (see the second and third columns). However, due to the post-infall metallicity gradients discussed in Section \ref{sec:gradient}, the high-energy tail of the Au-9-M debris is comprised of the least chemically evolved stars. Consequently, these metallicity differences are minimised in the regions where the debris of Au-9-M and its satellites overlap. This is a natural result, given that the metal-poor outskirts in these GSE-type mergers incorporate the debris of recently disrupted satellites (see the merger tree visualisations in Appendix \ref{appendix:mergertrees}). \par

In the fourth panel, we colour the debris by the vertical action ($J_z$). This can be interpreted as the the vertical excursion of particle orbits with respect to the plane of the galaxy \citep[see][for a mathematical description]{2008binney}. Actions are commonly used to assist in the identification of debris from various merger events \citep[e.g.][]{yuan2020b, limberg2021, malhan2022}. We calculate $J_z$ using {\sc agama} \citep{vasiliev2019} within a static axisymmetric potential fit to the mass of the central {\sc auriga} galaxy at $z=0$. The debris from both Au-9-M and its satellites span a wide range in $J_z$, with this variation being mostly independent of the boundaries between each debris group. An underlying pattern emerges, with $J_z$ at higher $L_z$ adopting the lowest values ($J_z < 10^2\,\rm{kpc}\,\rm{km}\,\rm{s}^{-1}$ for regions close to the perimeters of the $L_z$ distribution), and $J_z$ at higher energies adopting the highest values ($J_z > 10^3\,\rm{kpc}\,\rm{km}\,\rm{s}^{-1}$ for energies $>-0.7\times10^5\,\rm{km}^2\,\rm{s}^{-2}$). This same pattern appears across the rest of the {\sc auriga} selection. The wide range of $J_z$ values between each debris group, and even internal to larger debris groups, would make it difficult to draw connections between each debris. Some regions of the GSE-type debris may even appear to be unrelated to one another \citep[as in][]{amarante2022}. \par

% Satellite orbits
%%%%%%%%%%%%%%%%%%%%%%%%%%%%%%%%%%%%%%%%%%%%%%%%%%%%%%%%%%%%%%%%%%%%%%
\subsubsection{Surviving satellite remnants}

\begin{figure*}
\centering
  \setlength\tabcolsep{2pt}%
    \includegraphics[keepaspectratio, trim={0.0cm 0.0cm 0.0cm 0.0cm}, width=\linewidth]{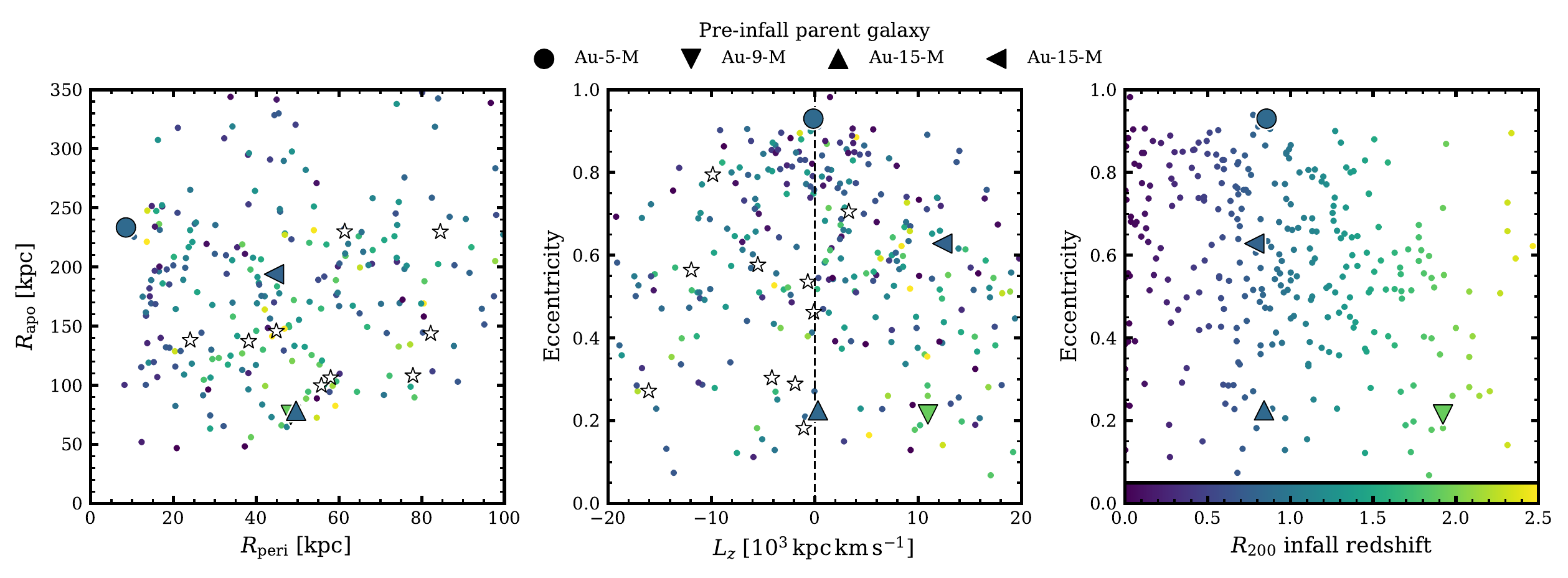}\\
\caption{Various orbital properties for surviving satellites in the dwarf-mass regime (defined here as $10^5 < M_\star/\text{M}_{\odot} < 10^7$). \textit{Left panel:} The apo- and pericentres for all satellites. \textit{Middle panel:} The orbital eccentricity versus the $z$-component of the specific angular momentum. \textit{Right panel:} The orbital eccentricity versus the infall redshift (simulation data only). White stars represent observed MW dwarf spheroidal galaxies from \citet{pace2022}. Small circles represent surviving satellites across the full {\sc auriga} suite, and are coloured by their infall time. The satellites which were associated with GSE-type mergers are shown with unique markers. These satellites are not exceptional in any obvious way, though this cannot be said with high confidence given that there are only four data points.}
\label{fig:sat_orbits}
\end{figure*}

Only a small fraction of GSE-type satellites survive until $z=0$, and only four of those are well-resolved ($M_{\star}>10^5\,\rm{M}_{\odot}$). However, the survival of these four satellites raises the possibility that one or more of the MW dwarf spheroidal galaxies may have originated from a GSE merger event. In this section, we ask whether it would be possible to distinguish these galaxies from the rest of the satellite population. \par

We show a selection of orbital properties for surviving satellites in the stellar mass range $10^5<M_\star/\rm{M}_{\odot}<10^7$ at $z=0$ in Figure \ref{fig:sat_orbits}. Satellites that are unaffiliated with GSE-type mergers are represented by points, where the colour corresponds to the infall redshift. The five satellites that were associated with GSE-type merger events are represented by enlarged, unique markers. In order to provide some context, we also include the properties of MW dwarf spheroidal galaxies from \citet{pace2022} as white stars, using the same stellar mass cuts. \par

For the simulation data, we select all satellites within the $R_{200}$ virial radius of the central {\sc auriga} galaxy at $z=0$. The infall redshift is defined as the last time the satellite crossed into this $R_{200}$ virial radius. The instantaneous orbital properties are then determined by integration within {\sc agama} \citep{vasiliev2019}, assuming a static and axisymmetric host density profile that was fit to the total mass of the central {\sc auriga} galaxy. For the observational data from \citet{pace2022}, we use the `edr3' values for the proper motion in Ra and Dec. These are then converted into galacto-centric coordinates using {\sc skycoord} from the {\sc astropy} package \citep{astropy}. We estimate the stellar mass from the $V$-band luminosity with a stellar mass-to-light ratio of $M/L_{V}=2$, which is a reasonable assumption for the older metal-poor stars in galaxies of this mass scale \citep{mclaughlin2005}. \par

There is no strong signature feature in the orbital properties of the surviving GSE-type satellites in Figure \ref{fig:sat_orbits}, and neither are there tight correlations in the full satellite populations. There is only a weak relationship between orbital eccentricity and infall redshift, as shown in the right panel, whereby an earlier infall leads to less eccentric orbits. This trend is impacted by survivorship bias; satellites on eccentric infalls are more likely to be disrupted by strong gravitational tides at the galactic centre. However, it is worth considering that the long-term survivability of these satellites is limited by the simulation resolution \citep[see][]{grand2021}. \par

% Stacked contributions
%%%%%%%%%%%%%%%%%%%%%%%%%%%%%%%%%%%%%%%%%%%%%%%%%%%%%%%%%%%%%%%%%%%%%%
\subsection{Implications of GSE-type mergers on the search for ancient disrupted relic galaxies} \label{sec:debristrends}

Debris linked to the GSE merger dominates the mass fraction of the stellar halo around the Solar neighbourhood \citep[e.g.][]{lancaster2019, myeong2022}. The same is true for many of the GSE-type mergers in {\sc auriga}. The overwhelming contribution from the GSE-type merger could be saturating the chemodynamical parameter space, thereby obscuring the signatures of other less massive mergers. \par

\subsubsection{Debris in the Solar neighbourhood and beyond} \label{GSE-debris p1}

\begin{figure*}
\centering
  \setlength\tabcolsep{2pt}%
    \includegraphics[keepaspectratio, trim={0.0cm 0.0cm 0.0cm 0.0cm}, width=\linewidth]{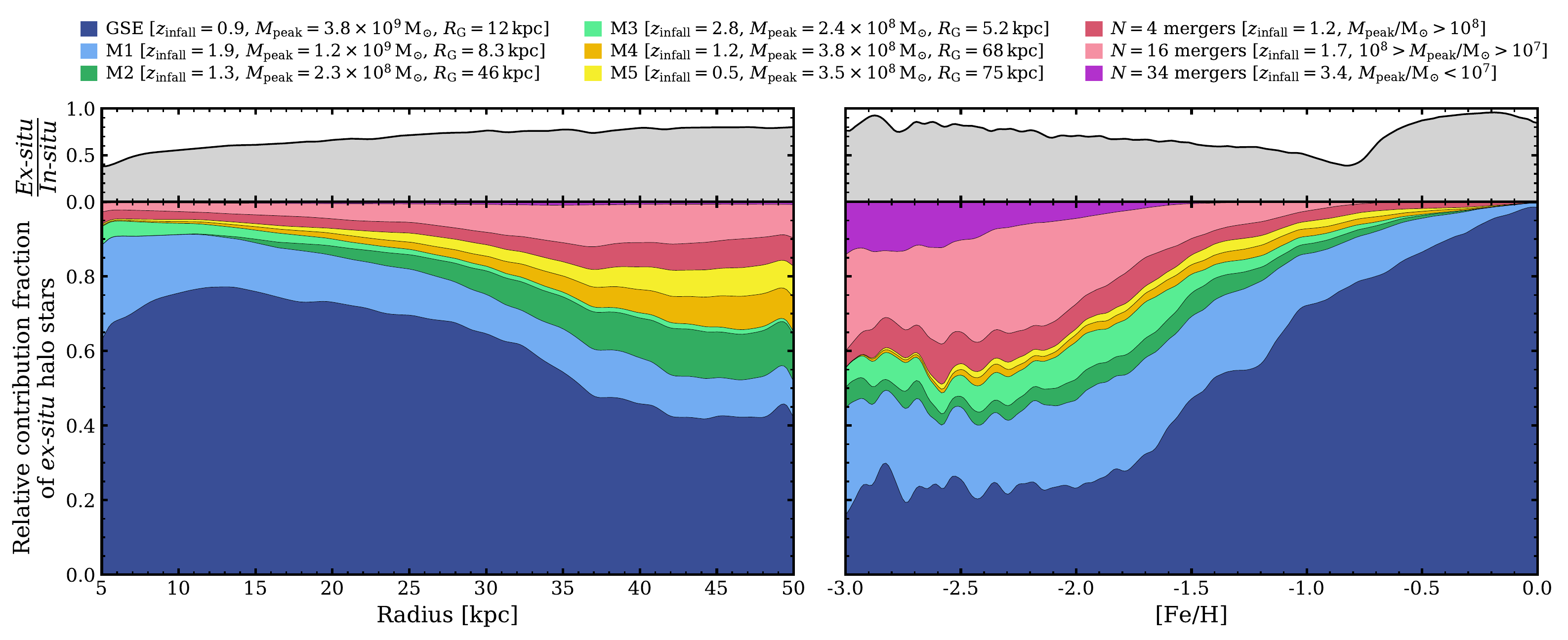}\\
\caption{The radial distributions (left panels) and metallicity distributions (right panels) of \textit{ex-situ} stars in the radial range $5<R_{\rm G}/\rm{kpc}<50$ in Au-5. `\textit{Ex-situ}' refers to stars that were originally unbound to the central {\sc auriga} galaxy. \textit{Upper panels:} A histogram of the \textit{ex-situ} to \textit{in-situ} stellar halo mass fraction, where disc stars have first been removed following the kinematic decomposition described in Section \ref{sec:post}. \textit{Lower panels:} Stacked histograms representing the contribution of different merger events, as a fraction of all \textit{ex-situ} stars. The GSE-type merger is labelled `Au-5-M', and then the five next most dominant mergers in this radial range are labelled `M1-5'. All other debris is binned according to the peak stellar mass of their progenitor galaxies. The histograms are constructed with a Gaussian KDE using Silverman's rule. In the case of individual mergers, the infall redshift, peak stellar mass, and the median orbital radius of the debris are included in the legend. In the case of mergers binned by their progenitor mass, the total number of distinct galaxies within this radial range are included in the legend, along with their median infall redshift. The relative fraction of GSE-type stars is greater than 50 per cent over most radii, but this fraction is reduced for increasingly low-metallicity stars. Similar plots for the other Auriga realisations are included at the following \href{https://drive.google.com/drive/folders/1teFkVLXATqKP3iYpDl8o0UnYHTtEjqJo?usp=sharing}{link}, or in the supplementary material.}
\label{fig:contributions_halo}
\end{figure*}

\begin{figure*}
\centering
  \setlength\tabcolsep{2pt}%
    \includegraphics[keepaspectratio, trim={0.0cm 0.0cm 0.0cm 0.0cm}, width=\linewidth]{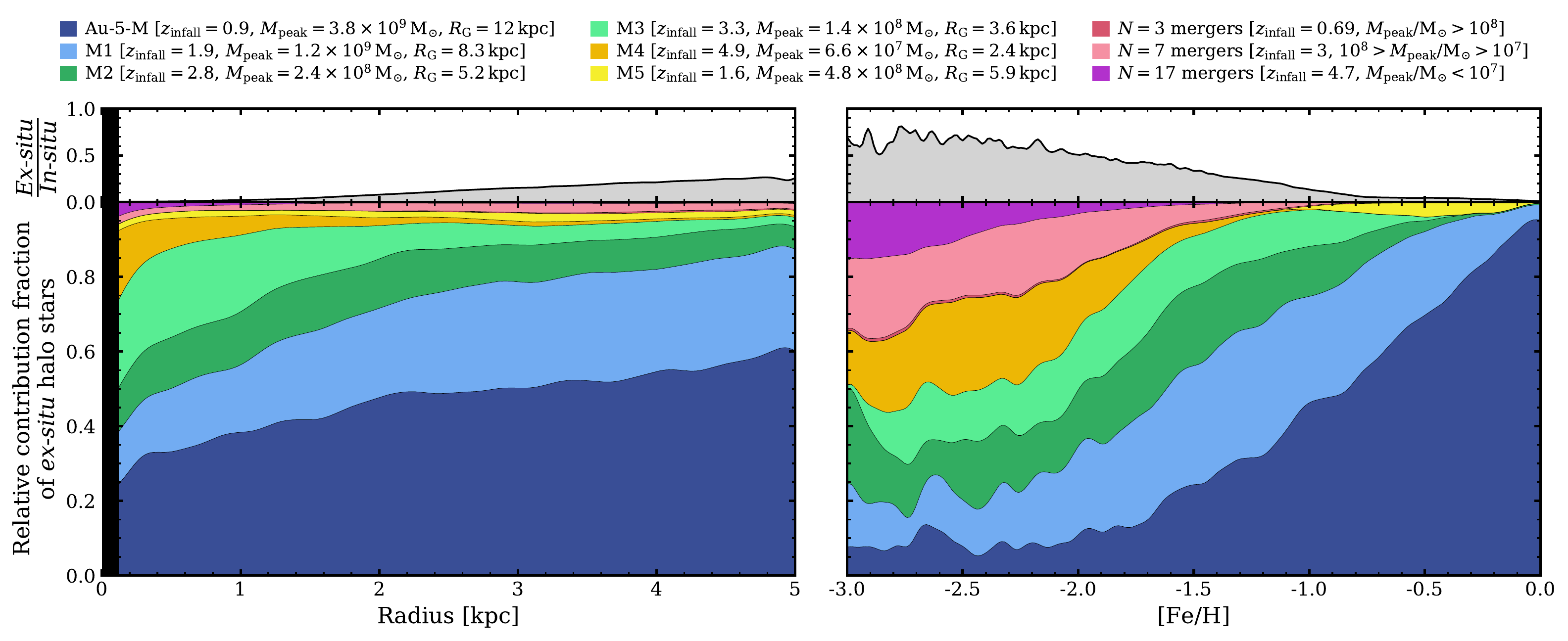}\\
\caption{The same as Figure \ref{fig:contributions_halo}, but for stars in an inner radial range of $0>R_{\rm G}/\rm{kpc}>5$. Regions with insufficient particle counts are filled in black. The stellar disc and bulge dominate within this regime ($R_{\rm d}=3.58\,\rm{kpc}$, $R_{\rm eff}=0.84\,\rm{kpc}$, \citealp{Auriga}). The five most dominant mergers in this inner radial range are not necessarily the same as in Figure \ref{fig:contributions_halo}, and typically sample older accretion events. As compared to Figure \ref{fig:contributions_halo}, the relative contribution fraction of the GSE-type merger is reduced. Similar plots for the other Auriga realisations are included at the following \href{https://drive.google.com/drive/folders/1teFkVLXATqKP3iYpDl8o0UnYHTtEjqJo?usp=sharing}{link}, or in the supplementary material.}
\label{fig:contributions_bulge}
\end{figure*}

\begin{figure*}
\centering
  \setlength\tabcolsep{2pt}%
    \includegraphics[keepaspectratio, trim={0.0cm 0.0cm 0.0cm 0.0cm}, width=\linewidth]{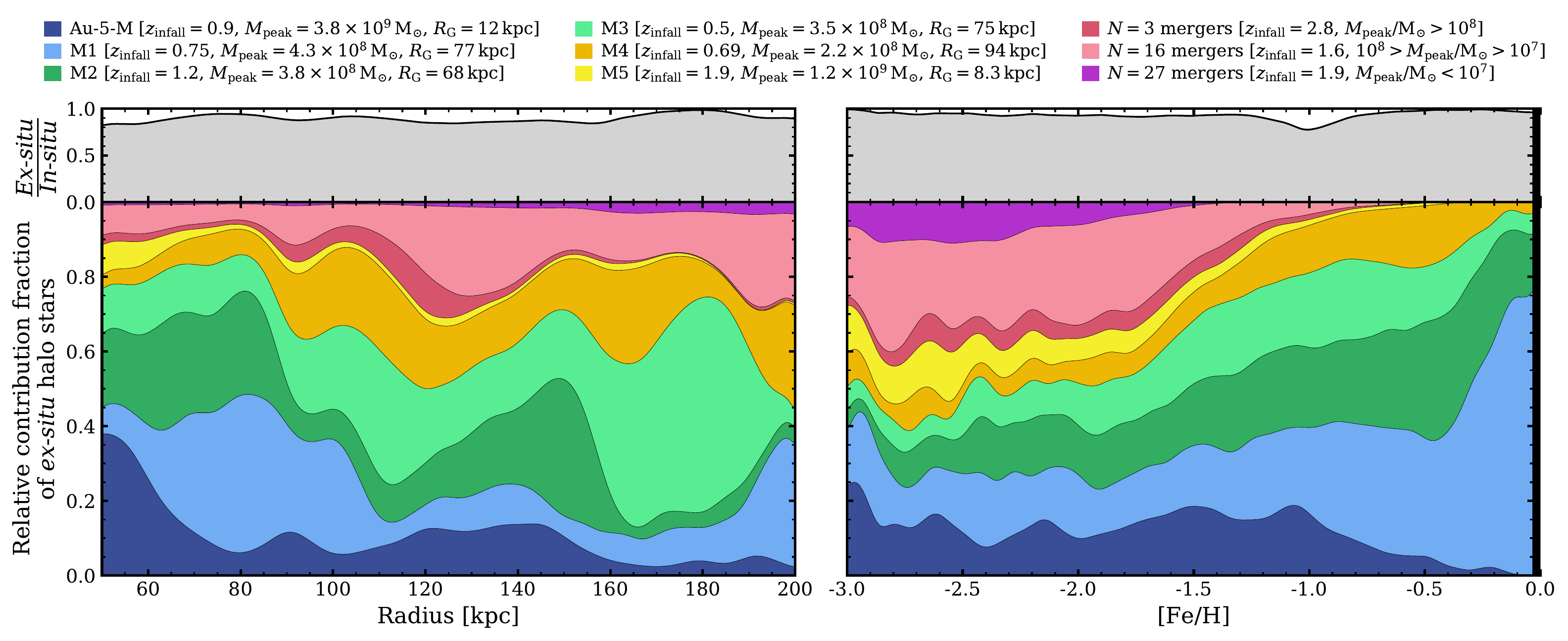}\\
\caption{The same as Figure \ref{fig:contributions_halo}, but for stars in an outer radial range of $50>R_{\rm G}/\rm{kpc}>200$. Regions with insufficient particle counts are filled in black. This regime is dominated by \textit{ex-situ} stars, originating from a wide variety of different accretion events. Again, the five most dominant mergers are not necessarily the same as in Figure \ref{fig:contributions_halo}. The debris from more recent mergers is favoured. Similar plots for the other Auriga realisations are included at the following \href{https://drive.google.com/drive/folders/1teFkVLXATqKP3iYpDl8o0UnYHTtEjqJo?usp=sharing}{link}, or in the supplementary material.}
\label{fig:contributions_outskirt}
\end{figure*}

Here, we investigate the relative fraction of GSE-type debris in one example simulation. We select Au-5, because its debris features are especially clear and the radially anisotropic debris feature at $z=0$ is heavily dominated by a single merger progenitor (see Section \ref{sec:contributions}). Furthermore, this realisation was shown to be the most comparable to the GSE in F19. Nonetheless, we note that the trends reported in this example are general across our {\sc auriga} selection, and we show this later in Section \ref{GSE-debris p2}. We also include the full plots for each realisation at the following \href{https://drive.google.com/drive/folders/1teFkVLXATqKP3iYpDl8o0UnYHTtEjqJo?usp=sharing}{link}, and see also the supplementary material. \par

Inspired by the halo decomposition exercise performed in \citet{naidu2020}, we dissect the \textit{ex-situ} halo of Au-5 in Figure \ref{fig:contributions_halo}. The stacked coloured histograms in the lower panels represent the contributions from different merger events to the fraction of all \textit{ex-situ} halo stars at $z=0$. We have excluded stars that remain bound to substructure. The grey histogram in the upper panels represents the fraction of \textit{ex-situ} to \textit{in-situ} halo stars. In all cases, co-rotating disc stars have been kinematically selected as in Section \ref{sec:post} and then removed. This cut includes \textit{ex-situ} stars that have adopted disc-like orbits. \par

The first coloured band, labelled ``Au-5-M'', is the stellar debris originating from the GSE-type merger. The bands labelled ``M1-5'' represent the next five mergers that contribute the largest fractions of the \textit{ex-situ} stellar mass over the radial range $5<R_{\rm G}/\rm{kpc}<50$. The final three histograms represent the remaining \textit{ex-situ} stars, binned with respect to the peak mass of their progenitor galaxies. \par

In the left panels, the debris is plotted with respect to its galacto-centric radius over the range $5<R_{\rm G}/\rm{kpc}<50$. The debris from the GSE-type merger dominates that of all other mergers, and contributes in excess of 50 per cent of all \textit{ex-situ} stars over the radii considered here. Stars from mergers with a progenitor stellar mass $M_\star<1\times10^8\,\rm{M}_{\odot}$ (the pink and purple bands) contribute a negligible fraction over all radii. \par

In the right panels, the debris is plotted with respect to its [Fe/H] metallicity over the range $-3 > \text{[Fe/H]} > 0$. We choose these limits because they effectively bracket the simulation data, whilst ensuring there are still a statistically meaningful number of star particles across the entire range. The fraction of GSE-type stars is almost 100 per cent around Solar metallicities ($\text{[Fe/H]} = 0$), but the relative abundance declines for lower metallicities ($\sim25$ per cent for $\text{[Fe/H]}<-2$). Stars from mergers with a progenitor mass $M_\star<1\times10^8\,\rm{M}_{\odot}$ now contribute nearly 40 per cent of \textit{ex-situ} stars at the lowest metallicities. Furthermore, the \textit{ex-situ} fraction in the stellar halo remains relatively high at these same low metallicities ($\sim90$ per cent). \par

This result can be explained by considering the histories of the mergers that contribute to the stellar halo at these radii. The GSE-type merger infalls at a relatively late epoch ($z=0.90$), and so it was free to evolve towards higher metallicities unimpeded. In contrast, the mergers marked ``M1-5'' infall at $z=3.33\mathit{\textnormal{-}}1.60$. Their mass and chemical evolution was quenched prematurely, with no or few stars attaining Solar metallicities. This is a natural consequence of GSE-type features typically owing to the last major merger that contributed to the inner galaxy. \par

All galaxies forming in relative isolation will begin their evolution from a low metallicity, and so there will be a fraction of early low-metallicity stars within the debris of every merger. The more massive mergers, such as GSE-type mergers, rapidly self-enrich and form a larger quantity of stars with raised metallicity. At the lowest metallicities, however, the relative contribution fraction is divided more equitably across all accreting satellites. \par

\subsubsection{Debris in the galactic centre}

With the continued improvements in observational data and analytical techniques, there is a growing interest in investigating the centre of the MW \citep[i.e.][]{ness2013, howes2017, lucey2019, arentsen2020, rix2022} -- a region that is often avoided due to dust contamination and crowding effects. The Galactic centre is likely to have been constructed in part by accretion events in the very early Universe, and is an ideal environment to search for ancient and low-metallicity merger debris \citep{schlaufman2014, badry2018, rix2022}. However, the short relaxation times and the phase-space bottleneck in the inner Galaxy rapidly erodes the rich kinematic landscape that is seen at the Solar radius and beyond \citep[e.g.][]{caterpillar}. This makes it far more challenging to identify and distinguish different accreted populations. \par

We reproduce our stacked histogram figures for the inner $5\,\rm{kpc}$ of Au-5 in Figure \ref{fig:contributions_bulge}. The contribution fraction of the GSE-type merger now descends below 50 per cent within the inner 2.5\,kpc, and the contributions of more ancient mergers are enhanced. When examining the metallicity distribution, a similar trend is seen in these inner regions as in the outer regions. However, the fractional contribution of low-mass mergers is proportionally greater. \par

The fraction of \textit{ex-situ} stars falls rapidly towards the galactic centre due to the presence of the \textit{in-situ} bulge and proto-halo \citep{gargiulo2019, fragkoudi2020, grand2020}, but most of these \textit{in-situ} stars are formed with higher metallicities. The \textit{ex-situ} fraction rises to around 60 per cent for metallicities of $\text{[Fe/H]}<-2$, and further to around 70 per cent for metallicities of $\text{[Fe/H]}<-2.5$. We anticipate these trends will continue to even lower metallicities, but we avoid commenting on this regime due to the reduced number of stellar particles and because the sub-grid physics models do not model the first generations of stars. \par

These results are encouraging in that even the most rudimentary parameter cuts can eliminate the majority of \textit{in-situ} and GSE-type contaminants. \par

\subsubsection{Debris in the galactic outskirts}

The outskirts of the MW stellar halo are sparsely populated with few confirmed stars. Whilst it is currently prohibitive to detect more than the brightest populations (e.g. blue horizontal branch \citealp{deason2012}, M-giant stars \citealp{Bochanski2014} and RR Lyrae \citealp{cohen2017}), these outskirts will be revealed in ever-increasing detail by future observational surveys (e.g. JWST \citealp{gardner2006}, LSST \citealp{LSST2009}, WFIRST \citealp{WFIRST2015}). \par
% see https://arxiv.org/pdf/1609.06406.pdf

We reproduce our stacked histogram once again, this time for the range $50<R_{\rm G}/\rm{kpc}<200$ in Figure \ref{fig:contributions_outskirt}. We do not extend this figure out to the full virial radius, because the number of star particles is too low for statistically meaningful analysis. Predictably, there are very few \textit{in-situ} stars present. \par

The contribution fraction of the GSE-type debris is far less significant at these higher radii, and this is because there is a greater contribution from an assortment of other merger debris. The mergers that contribute to this radial regime tend to have been accreted more recently than those at lower radii, and there is a greater proportion of stars from low-mass mergers (pink and purple bands) which were rapidly disrupted upon infall. This is entirely expected and in accordance with previous works \citep[e.g.][]{bullock2005, fattahi2020}. \par

The metal-rich core of the GSE-type merger is shielded against tidal disruption until it has fallen to within $\sim50\,$kpc of the central {\sc auriga} galaxy. Consequently, it contributes almost 0 per cent of stars approaching Solar metallicities. Instead, the contribution fraction becomes dominated by metal-rich mergers that were disrupted at higher radii. This behaviour is not seen in Au-10 or Au-18, because those galaxies do not undergo such mergers. \par

\subsubsection{Overall trends} \label{GSE-debris p2}

\begin{figure*}
\centering
  \setlength\tabcolsep{2pt}%
    \includegraphics[keepaspectratio, trim={0.0cm 0.0cm 0.0cm 0.0cm}, width=0.7\linewidth]{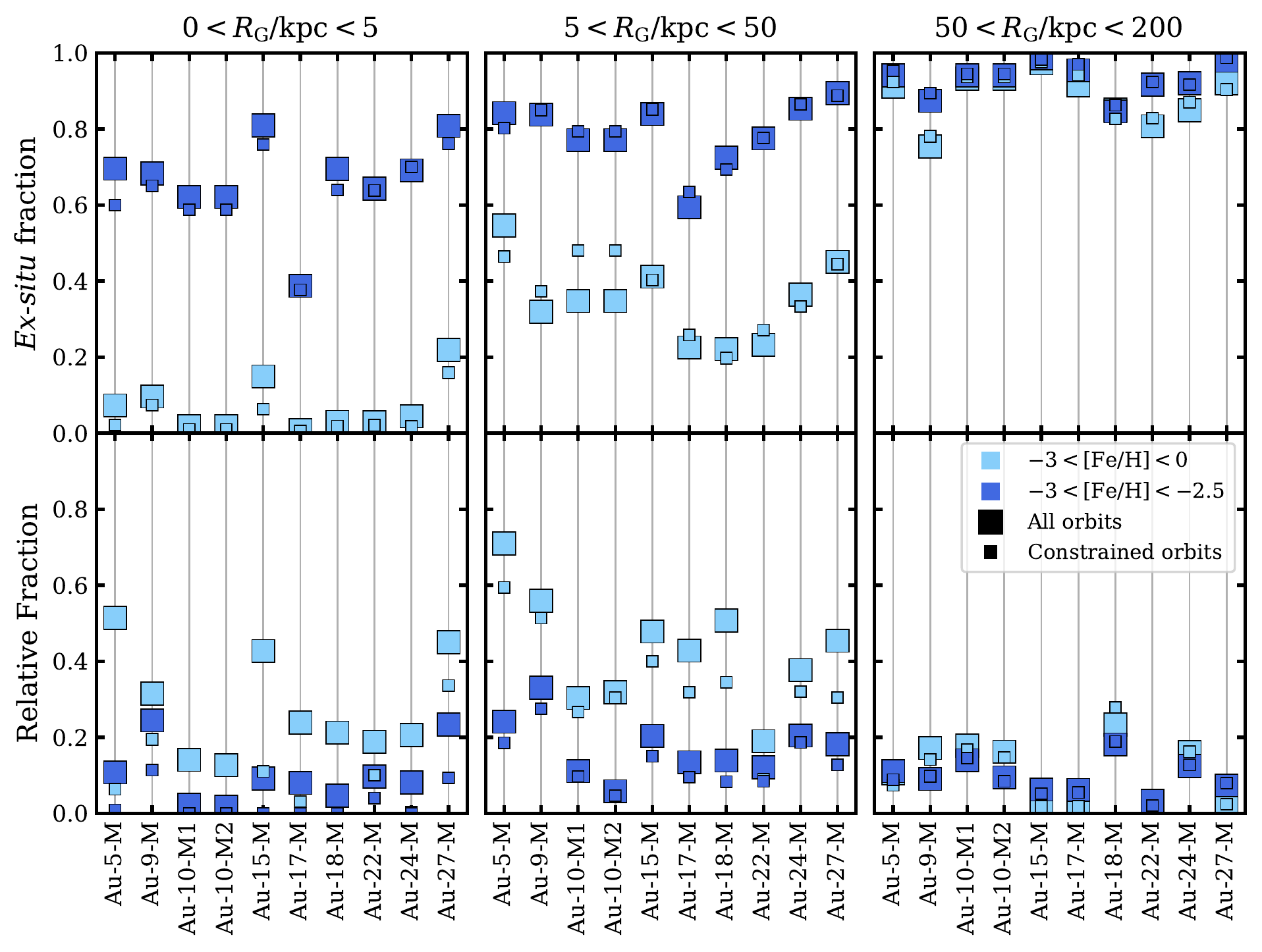}\\
\caption{\textit{Upper panels}: The fraction of \textit{ex-situ} to \textit{in-situ} halo stars (excluding the stellar disc). \textit{Lower panels}: The fraction of \textit{ex-situ} stars that originated from a GSE-type merger event. The panel titles indicate the radial limits considered. The marker colour represents different metallicity cuts, with all stars shown in light blue and metal-poor stars in dark blue. The marker size represents different orbital cuts, where all stars are shown with large markers and orbits constrained within the radial limits are shown with small markers. In these cases, all stellar pericentres are greater than the minimum radius and all stellar apocentres are less than the maximum radius.}
\label{fig:cake_fractions}
\end{figure*}

We illustrate trends across the rest of our {\sc auriga} selection in Figure \ref{fig:cake_fractions}. The upper panels represent the \textit{ex-situ} to \textit{in-situ} stellar halo fractions, and the lower panels represent the fraction of \textit{ex-situ} stars that originated from GSE-type mergers. The GSE-type debris in Au-10 is shown for each merger individually, though they could arguably be considered as a single debris population. The left column represent stars in the range $0<R_{\rm G}/\rm{kpc}<5$ as in Figure \ref{fig:contributions_bulge}, the middle column represent stars in the range $5<R_{\rm G}/\rm{kpc}<50$ as in Figure \ref{fig:contributions_halo}, and the right column represent stars in the range $50<R_{\rm G}/\rm{kpc}<200$ as in Figure \ref{fig:contributions_outskirt}. We perform cuts on the metallicity in order to illustrate the difference between a wide range of metallicities ($-3<\text{[Fe/H]}<0$) and metal-poor stars ($-3<\text{[Fe/H]}<-2.5$). \par

Our basic radial cuts do not account for stars that are on orbits which take them outside of the defined radial limits. We estimate the peri- and apocentres for each star particle by integrating its full orbit in an axisymmetric potential using {\sc agama} \citep{vasiliev2019}. We show all stars using large markers, and then stars that remain within the stated radial range over their entire orbits using small markers. \par

The figure shows that the \textit{ex-situ} fraction is far lower in the inner radial range, and this is due to the presence of dense \textit{in-situ} stellar bulges within the inner few kpc. However, the \textit{ex-situ} fraction is increased when metal-rich stars are excluded. \par

In almost all cases, the relative fraction of GSE-type debris is reduced once more metal-rich stars are excluded. As mentioned in Section \ref{GSE-debris p1}, this is because the bulk of the stars in these GSE-type mergers are more enriched than the rest of the \textit{ex-situ} population, which is a direct consequence of the GSE-type mergers typically being among the most recent and massive mergers that contributed to the inner galaxy. \par

The effect of the orbital cuts is subdominant compared to the metallicity cuts, and with predictable outcomes. The \textit{ex-situ} fractions over the radii $5<R_{\rm G}/\rm{kpc}<50$ are, on average, increased after the orbital cuts. This is because the cuts remove high eccentricity \textit{in-situ} stars from the dense bulge region, which have pericentres that take them below the 5\,kpc lower limit. In contrast, the \textit{ex-situ} fraction in at $0<R_{\rm G}/\rm{kpc}<5$ are slightly reduced after the orbital cuts. This is because there are many \textit{ex-situ} stars on highly radial orbits, such as those from the GSE-type mergers, which have apocentres greater than 5\,kpc. It is for this same reason that the relative fraction of GSE-type stars is, on average, reduced after the orbital cut. \par

There is little difference between different parameter cuts for radii $50<R_{\rm G}/\rm{kpc}<200$, with uniformly low fractions of stellar debris from GSE-type mergers. \par

In summary, these results indicate that the contamination from GSE-type debris is reduced at lower metallicities (e.g. $\text{[Fe/H]}<-2.5$), towards the galactic centre (e.g. $R_{\rm G}<5\,\rm{kpc}$) and outskirts (e.g. $R_{\rm G}>50\,\rm{kpc}$), and excluding stars with higher orbital apocentres (e.g. $r_{\rm apo} > 5\,\rm{kpc}$). More specifically, both the fraction of GSE-type debris and the fraction of \textit{in-situ} stars are reduced for increasingly low metallicity. It is possible that this trend continues even below $\text{[Fe/H]}=-3$, although there are few star particles below this limit and so we avoid making any firm interpretations. We also note that the fraction of \textit{in-situ} stars can only ever be overestimated due to the manner in which they are identified, meaning the \textit{ex-situ} fractions reported here should be considered as a lower bound. \par

In Appendix \ref{appendix:contributions_contour}, we test whether these relationships fail for stars that occupy similar $E/L_z$ as the GSE-type debris. We find that it makes little difference. \par

\section{Discussion} \label{sec:discussion}

\subsection{Diversity and similarities of GSE-type mergers}

The GSE-type mergers presented in Table \ref{tab:gse_analogue_properties} span a wide range of progenitor properties, many of which have been discussed in this work. In some cases, there is more than one single merger which contributes a large mass fraction to the radially anisotropic stellar debris near the Solar radius ($\sim 8\,\rm{kpc}$). \par

Despite this inherent diversity, there is a great deal of degeneracy in the final debris footprints. As described in F19, the resulting GSE-type features contain a dominant component with velocity anisotropy of $\beta>0.8$, and a contribution in excess of 50 per cent to the stellar halo. Furthermore, the overall form of the debris in energy-space and configuration-space converge upon similar qualitative properties. \par

There is a great stochasticity in both the number and mass of pre-infall luminous satellites associated with GSE-type mergers, with little dependence on the stellar mass of the host. Furthermore, the majority of these satellites are separated from their parent GSE-type merger soon after infall, and do not descend to the low specific orbital energies that is typical for GSE-type debris. Only a marginal fraction of these satellites survive as remnant objects at $z=0$, though we note that their survival may be impacted by resolution effects \citep[see][]{grand2021}. These few survivors do not have exceptional orbital properties that might indicate that they were once related to the GSE-type merger event. \par

We show that the debris from the superposition of two separate mergers may be almost entirely degenerate with one another. This is a similar problem to that raised in \citet{jeanbaptiste2017}, where it is found that the debris from various merger events can overlap and fragment in ways that make it difficult to determine their origins. Whilst ages and chemical abundances can alleviate those challenges, they are less helpful in the case of two near-mass mergers that merge at a similar time. This could make it nearly impossible to confirm whether a GSE-like debris feature was sourced from one or two progenitors, especially since such a scenario would only change the progenitor mass estimate by a factor of two -- which is already well within the range of current estimates for the GSE. Similar results were found in \citet{rey2023}, where it is shown that a $\Lambda$CDM cosmology naturally leads to several mergers contributing radially anisotropic halo debris, and that the properties of this debris are relatively insensitive to its assembly. There are also arguments for this `multiple radial merger' scenario based on observational data of halo stars \citep{donlon2022, donlon2023}. \par

In essence, it is exceedingly difficult to accurately constrain many of the progenitor properties from the chemodynamics of merger debris at $z=0$, even when the privileges of simulation data are readily available. \par

\subsection{Revealing the centre of the MW}

% Introduce the concept:
In the $\Lambda$CDM cosmology and hierarchical galaxy formation more generally, the proto-MW is assembled from a spectrum of high mass-ratio major mergers at early times \citep[e.g.][]{renaud2021, renaud2021b}. Much of the debris from these ancient mergers will remain locked in the low-energy potential well, concealed amongst the stellar bulge, bar and disc \citep[e.g.][]{bullock2005, wetzel2011, rocha2012, bosch2016, starkenburg2017}. \par

% Evidence for central merger debris from GCs:
There has been growing evidence for this scenario in the MW. Analysis of the age and chemical distribution of globular cluster populations suggest they were either born \textit{in-situ} or donated by known satellite accretions \citep{Kruijssen2019, massari2019, forbes2020}. Some 16 per cent, however, are associated with an unknown group at low binding energies. This led to the prediction of a high mass-ratio ancient merger, accreting around $z=1$ with a progenitor stellar mass in the region $\sim2\times10^9\,\rm{M}_{\odot}$ \citep{Kruijssen2019}. This merger was named \textit{Kraken}, and possible evidence of its debris have since been uncovered near the bottom of the MW potential well in \textit{Gaia} and APOGEE data \citep{horta2021, naidu2022}. \par

% Kraken might not be a true merger event:
Further examination has suggested that \textit{Kraken} may not be a genuine merger at all, but rather a low-energy extension to the GSE debris or even a misidentified \textit{in-situ} population. \citet{lane2022} highlighted that the apparently distinct energy level of \textit{Kraken} debris was aligned with an energy gap in the survey selection function of APOGEE data. However, the distinct chemistry of \textit{Kraken} stars cannot easily be explained if they were exclusively an extension of the low-energy GSE tail (i.e. \citealp{amarante2022}). Around the same time, \citet{belokurov2022} identified a population of old, high-$\alpha$ and \textit{in-situ} stars toward the MW bulge which they named \textit{Aurora}. The chemistry of this population overlaps with the \textit{Kraken} debris, and the high-apocentre tail of \textit{Aurora} stars could feasibly travel out of the bulge and masquerade as \textit{Kraken} stars. Further decomposition of the stellar halo have supported this scenario \citep{myeong2022, rix2022}, and \citet{orkney2022} showed that debris from a \textit{Kraken}-style merger may be difficult to distinguish from contemporaneous \textit{in-situ} populations. On the other hand, \citet{horta2022} are able to separate the \textit{Aurora} and \textit{Kraken} populations by their distinct $\alpha$-abundance at low metallicity using a promising statistical method. It would, therefore, be incredibly valuable to assess the significance of this result using mock data. \par

% Conclude this intro:
It is clear that the pervasiveness of the GSE debris, and contamination from other stellar populations, are a real danger when investigating the inner MW. Nonetheless, the origins of those remaining globular cluster populations remains to be conclusively decided, which could suggest yet undiscovered high-redshift merger debris. \par

% Report/discuss the result:
We have shown that, across our simulation sample, debris from the main GSE-type progenitor is the single largest contributor to the stellar halo around the Solar neighbourhood, making up roughly 50 per cent of all \textit{ex-situ} halo stars. However, this relative contribution is greatly reduced within the inner 5\,kpc and for metallicities $\text{[Fe/H]}<-2.5$. The contribution can be decreased even further by excluding stars with apocentres greater than 5\,kpc, in some cases to less than 5 per cent. The \textit{ex-situ} halo mass fraction is exceedingly low in the inner few kpc (typically less than 20 per cent, with much of the contamination arising from kinematically heated disc stars). However, this fraction is raised dramatically when considering only the most metal-poor stars (to around 70 per cent in most cases). \par

From a philosophical point of view, it becomes increasingly difficult to consider any single progenitor galaxy to be the \textit{main} MW progenitor before a time of $\sim z=6$. Many of the different progenitor components would be of a similar mass and star formation rate at these earliest times (see for example Appendix \ref{appendix:mergertrees}), and so the distinction between \textit{in-} and \textit{ex-situ} becomes more and more meaningless. It can be argued that all stars formed within the deepest potential wells at $\sim z>6$ should be considered \textit{in-situ}, regardless of whether any of those potential wells were the \textit{main} progenitor or not \citep[e.g.][]{rix2022}. Alternatively, these early stellar populations could be distinguished based on whether they are chemically `evolved' or `unevolved' \citep[e.g.][]{fernandes2022}. Whatever their label, the stars from these earliest mergers offer a privileged view into processes that governed the early Universe, and so their detection is of paramount importance. \par

% Discuss the observational context and future surveys:
Investigating the centre of the MW incurs numerous challenges. There is high stellar crowding, obscuration from foreground populations and interstellar dust, ongoing active star formation, and low-metallicity stars are exceedingly rare. However, there are also distinct advantages. A large number of stars can be investigated with relatively few pointings and less volume coverage. The debris from ancient disrupted dwarfs will have lower orbital apocentres that are contained within the Galactic centre, meaning that low-metallicity stars identified here are more likely to be genuine ancient debris, as opposed to at the Solar radius and beyond where there are visitations from a wide range of eccentric orbits. Moreover, in the search for the very first stars, the Galactic centre is far closer and more available than other likely environments -- such as distant dwarf galaxies. \par

Observational surveys are beginning to resolve the inner MW in greater and greater detail (e.g. ARGOS \citealp{ness2013}, EMBLA \citealp{howes2017}, COMBS \citealp{lucey2019}, PIGS \citealp{arentsen2020} and also APOGEE \citealp{rix2022}), with increasing focus on metal poor stars below $\text{[Fe/H]}=-2$. These surveys reveal a high stellar density in the central few kpc of the MW, with a metallicity-dependent rotation that vanishes at around $\text{[Fe/H]}\simeq-2$ \citep[i.e.][]{arentsen2020} -- possibly revealing a transition to a pressure-supported classical bulge \citep[e.g.][]{Babusiaux2010}. There is now evidence that some of these stars were not born \textit{in-situ}, and arrived via globular cluster or dwarf galaxy merger \citep{sestito2023}. The Galactic centre may therefore offer a resolved view of the stellar populations formed in pre-reionisation galaxies, which would be complementary to the insights provided by the JWST \citep{gardner2006}. This, combined with cutting-edge spectroscopic surveys (SDSS-V \citealp{sdss52017}, 4MOST \citealp{4MOST}), MOONS \citealp{MOONS}, and asteroseismic surveys focused on metal-poor and dense stellar fields (e.g. HAYDN \citealp{HAYDN}), promises an unprecedented understanding of the Galactic centre and the physics that govern the first galaxies. Therefore, it is encouraging that the vast debris from the GSE and its satellites could potentially be screened with relatively simple parameter cuts. \par

\section{Conclusions} \label{sec:conclusion}

We have investigated a selection of GSE-like merger events from nine MW-like galaxies in the {\sc auriga} simulation suite. \citet{fattahi2019} originally identified radially anisotropic debris features within these simulations, with properties comparable to that of the velocity ellipsoid observed in the MW \citep{belokurov2018}. They showed that the bulk of this debris can be traced back to progenitor galaxies that accreted onto their hosts over the range $2.9>z>0.75$, with peak stellar masses spanning roughly an order of magnitude ($3\times10^8<M_{\star}(\rm{peak})/\rm{M}_{\odot}<4\times10^9$). We list our core results below. \par

\begin{itemize}
\setlength{\itemindent}{0.5em}
    \item In six of the nine simulations, \citep{fattahi2019} show that the vast majority of the radially anisotropic stars (velocities of $|v_{\phi}|<50\,\rm{km}\,\rm{s}^{-1}$; $100<|v_{\rm r}|/\rm{km}\,\rm{s}^{-1}<400$) are associated with a single merger event. We find that there are some instances where a second or even a third merger contribute significant mass fractions of their own. Some of these lesser contributions are also radially anisotropic, and it is not always possible to distinguish them by their chemodynamic properties. Similar results were found independently in \citet{rey2023}.
    \item The GSE-type progenitor galaxies exhibit a wide range of properties, including differing metallicity gradients and both rotation and pressure supported systems. However, the final debris distributions do not strongly relate to the progenitor properties.
    \item The GSE-type debris is initially prolate and tilted with respect to the disc, but dynamical processes drive an evolution towards a round or oblate shape and close alignment with the disc. This is in tension with the prolate and inclined shape of GSE debris reported in \citet{iorio2019, han2022, lane2023}, and may indicate that the early stellar mass enhancement in the {\sc auriga} physics model leads to exaggerated torques between the disc and other galactic components.
    \item Many of the GSE-type mergers are accompanied by their own luminous satellite populations, with between $0\mathit{\textnormal{-}}8$ of stellar mass $>10^6\,\rm{M}_{\odot}$. The majority of these are fully disrupted by $z=0$, and in most cases are not dragged to low orbital energies along with the main merger debris. Of those few satellites that survive as remnants, there is no clear indication in their orbital properties that they were once associated with the GSE-type merger.
    \item The progenitor mergers of GSE-like debris features contribute high mass fractions to the stellar halo around the Solar neighbourhood, and in some cases out to larger radii. However, these fractions are in most cases reduced to below 20 per cent when considering the central few kpc and metallicities below $\text{[Fe/H]}<-2.5$. The fraction of stars from \textit{ex-situ} sources is simultaneously increased when considering these low metallicities, increasing from less than 10 per cent in the inner few kpc to around 70 per cent. This highlights the centre of the MW as a promising environment to search for the ancient stars that formed in pre-reionisation dwarf galaxies, and a convenient alternative to distant objects at high redshifts. 
\end{itemize}

\section*{Acknowledgements}
We thank João Amarante for his insightful comments and useful discussion.
We thank the anonymous referee for their constructive advice that has greatly improved this work.
CL \& MO acknowledge funding from the European Research Council (ERC) under the European Union’s Horizon 2020 research and innovation programme (grant agreement No. 852839).
RG acknowledges financial support from the Spanish Ministry of Science and Innovation (MICINN) through the Spanish State Research Agency, under the Severo Ochoa Program 2020-2023 (CEX2019-000920-S), and support from an STFC Ernest Rutherford Fellowship (ST/W003643/1).
FvdV is supported by a Royal Society University Research Fellowship.
FAG acknowledges support from ANID FONDECYT Regular 1211370 and by the ANID BASAL project FB210003. FAG, acknowledges funding from the Max Planck Society through a “Partner Group” grant.
AF is supported by a UKRI Future Leaders Fellowship (grant no. MR/T042362/1)
This work used the DiRAC@Durham facility managed by the Institute for Computational Cosmology on behalf of the STFC DiRAC HPC Facility (www.dirac.ac.uk). The equipment was funded by BEIS capital funding via STFC capital grants ST/K00042X/1, ST/P002293/1, ST/R002371/1 and ST/S002502/1, Durham University and STFC operations grant ST/R000832/1. DiRAC is part of the National e-Infrastructure.
%%%%%%%%%%%%%%%%%%%%%%%%%%%%%%%%%%%%%%%%%%%%%%%%%%
\section*{Data Availability}
 
Available upon request.



\bibliographystyle{mnras}
\bibliography{references} % if your bibtex file is called example.bib

%%%%%%%%%%%%%%%%%%%%%%%%%%%%%%%%%%%%%%%%%%%%%%%%%%

%%%%%%%%%%%%%%%%% APPENDICES %%%%%%%%%%%%%%%%%%%%%
\appendix

\section{Merger tree} \label{appendix:mergertrees}

\begin{figure*}
\centering
  \setlength\tabcolsep{2pt}%
    \includegraphics[keepaspectratio, trim={0.0cm 0.0cm 0.0cm 0.0cm}, width=\linewidth]{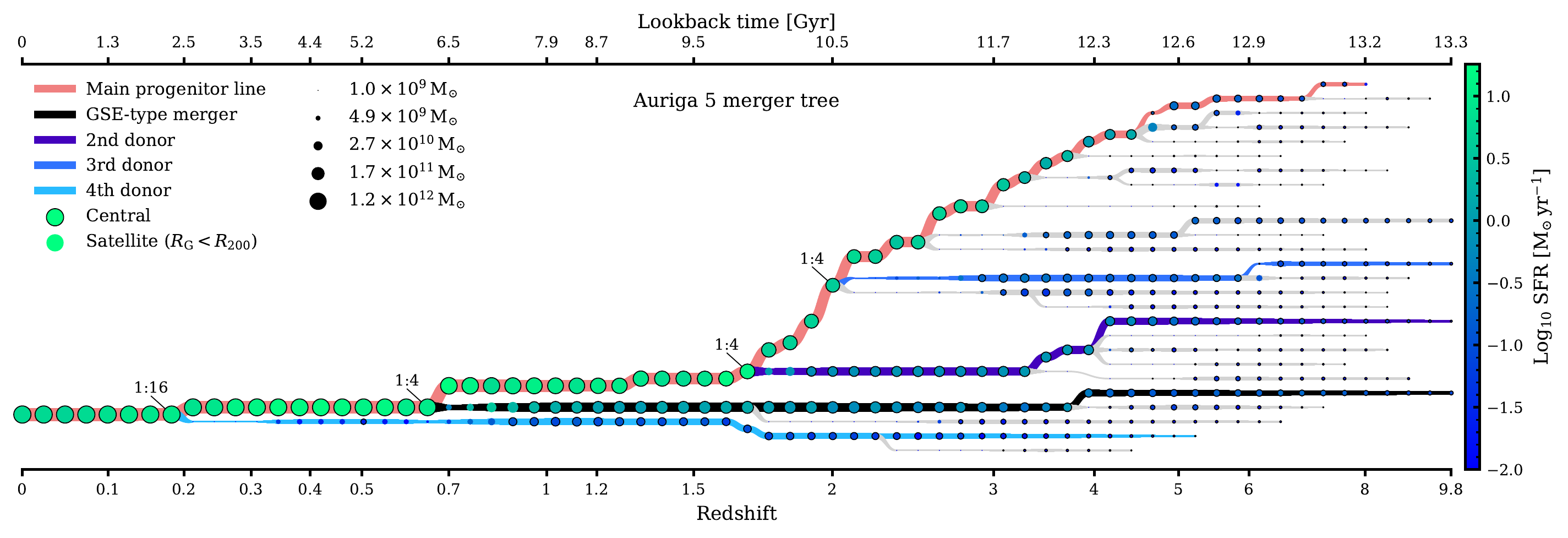}\\
\caption{A merger tree visualisation for Au-5. See main text for details.}
\label{fig:merger_tree}
\end{figure*}

Here we show a visualisation of the {\sc lhalotree} merger tree for Au-5 in Figure \ref{fig:merger_tree}. The formatting was made with the assistance of {\sc GraphViz} and the {\sc pydot} package for {\sc Python}. Each node represents a separate subhalo, with the connecting lines indicating the descendants (left) and progenitors (right). The node size represents the total mass of the subhalo as identified by {\sc subfind}. The node colour is the instantaneous star formation rate over all gas cells. We consider all mergers that are greater than $1:20$, and exclude haloes with no progenitors greater than a halo mass of $10^9\,\rm{M}_{\odot}$. \par

The main progenitor lines of the final halo, and of the notable mergers from Figure \ref{fig:paper_pie}, are shown with distinct colours as described in the legend. These notable mergers are annotated with a merger mass ratio, which describes the ratio of the $M_{200}$ virial mass between the main progenitor line and the merger progenitor line at a time before infall. \par

Merger tree visualisations for the other {\sc auriga} simulations presented in this paper are included at this  \href{https://drive.google.com/drive/folders/1teFkVLXATqKP3iYpDl8o0UnYHTtEjqJo?usp=sharing}{link}, and see also the supplementary material. \par

\section{Halo contributions within the GSE $E/L_z$ locus} \label{appendix:contributions_contour}

\begin{figure*}
\centering
  \setlength\tabcolsep{2pt}%
    \includegraphics[keepaspectratio, trim={0.0cm 0.0cm 0.0cm 0.0cm}, width=0.9\linewidth]{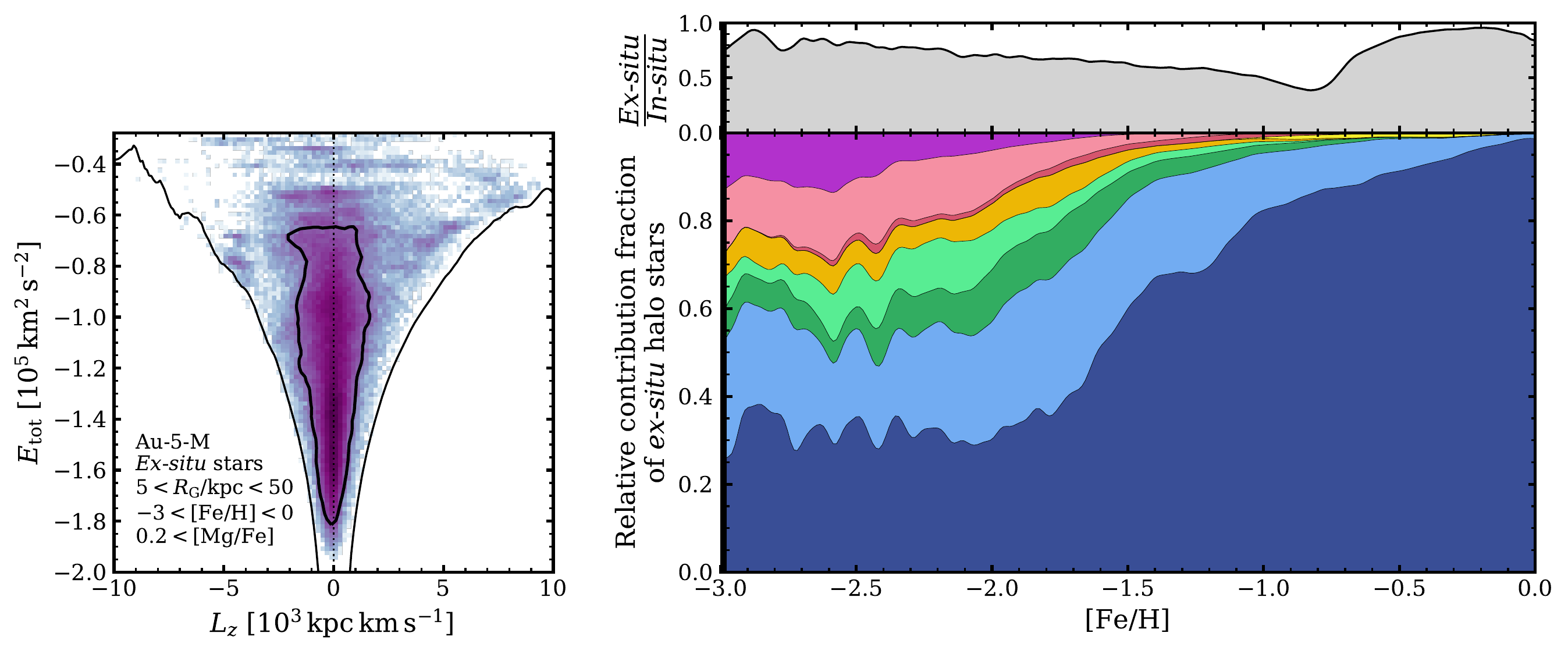}\\
\caption{\textit{Left panel:} A histogram of the \textit{ex-situ} halo stars in Au-5, as shown in $E/L_z$ parameter space. The parameter cuts are listed in the lower left corner. A thick black contour encircles 90 per cent of the mass associated with the GSE-type merger event. \textit{Right panels:} The same as in Figure \ref{fig:contributions_halo}, but for stars within the contour selection defined in the left panel. This shows that the proportion of stars from the GSE-type merger (dark blue band) are reduced for metallicities $\rm{[Fe/H]} < -2.5$, even within the $E/L_z$ region dominated by GSE-type debris.}
\label{fig:contributions_contour}
\end{figure*}

In Section \ref{sec:debris_fractions}, we argue that the contribution fraction of GSE-type merger debris is lowest for halo stars with metallicity $\rm{[Fe/H]} < -2.5$, and this suggests that other merger debris would be more effectively distinguished from GSE-type debris in the low-metallicity regime. It is conceivable that this trend might fail for debris that shares the same locus in $E/L_z$ as the GSE-type merger debris. \par

In Figure \ref{fig:contributions_contour} we reproduce the right panels of Figure \ref{fig:contributions_halo}, but for a selection in $E/L_z$ that encompasses 90 per cent of the GSE-type stellar debris mass. We include a chemical cut of $\rm{[Mg/Fe]} > 0.2$, intended to highlight merger debris associated with ancient dwarf galaxies. The left panel shows a histogram of all \textit{ex-situ} stars in $E/L_z$, where the thick black contour is the GSE-type debris selection. The right panels show the relative contribution of the stars within that contour and with the same parameter cuts, using the same colour scheme as in Section \ref{sec:debris_fractions}. \par

Even though the contour is specifically limited to the $E/L_z$ space that is dominated by GSE-type debris, there remains a large fraction of stars from other merger events. The GSE-type debris fraction at metallicities of $\rm{[Fe/H]} < -2.5$ is 30 per cent, up from 24 per cent when no $E/L_z$ selection is made. The results are similar across the rest of our {\sc auriga} selection, with the $E/L_z$ selection making little or no difference to the low fraction of GSE-type debris at low metallicity. \par

%%%%%%%%%%%%%%%%%%%%%%%%%%%%%%%%%%%%%%%%%%%%%%%%%%

% Don't change these lines
\bsp	% typesetting comment
\label{lastpage}
\end{document}